\documentclass[letterpaper,11pt]{article}
\pdfoutput=1

\usepackage{jheppub}

\usepackage{multirow}

\usepackage{subfig}
\usepackage{xspace}
\usepackage[countmax]{subfloat}

\allowdisplaybreaks

\usepackage{graphicx}
\usepackage{epstopdf}

\newcommand{\mut}{\tilde\mu}

\newcommand{\xlog}{{\rm W}}

\newcommand{\wideeq}[1]{\makebox[{#1}]{$=$}}

\newcommand{\as}{\alpha_s}
\newcommand{\order}[1]{{\cal O}\left(#1\right)}

\definecolor{darkblue}{rgb}{0,0,0.5}

\definecolor{darkgreen}{rgb}{0,0.7,0}

\definecolor{darkred}{rgb}{0.7,0,0}

\definecolor{grey}{rgb}{0.5,0.5,0.5}

\newcommand{\C}[2]{C^{(#2)}_{#1}}
\newcommand{\ECF}[2]{\text{ECF}\left(#1,#2\right)}

\newcommand{\zcut}{z_\text{cut}}
\newcommand{\ea}{{\C{1}{\alpha}}}

\DeclareRobustCommand{\Sec}[1]{Sec.~\ref{#1}}

\DeclareRobustCommand{\App}[1]{App.~\ref{#1}}

\DeclareRobustCommand{\Fig}[1]{Fig.~\ref{#1}}
\DeclareRobustCommand{\Figs}[2]{Figs.~\ref{#1} and \ref{#2}}

\DeclareRobustCommand{\Eq}[1]{Eq.~(\ref{#1})}

\DeclareRobustCommand{\Ref}[1]{Ref.~\cite{#1}}
\DeclareRobustCommand{\Refs}[1]{Refs.~\cite{#1}}

\newcommand{\pythia}[1]{\textsc{Pythia\xspace #1}}

\newcommand{\fastjet}[1]{\textsc{FastJet\xspace #1}}

\DeclareRobustCommand{\order}[1]{{\cal O}(#1)}

\preprint{ 
\begin{flushright}
DCPT/14/24\\
  IPPP/14/12 \\ MIT--CTP 4531
 \end{flushright}}

\title{Soft Drop}

\author[a]{Andrew J. Larkoski,}
\author[b]{Simone Marzani,}
\author[c]{Gregory Soyez,}
\author[a]{and Jesse Thaler}

\affiliation[a]{Center for Theoretical Physics, Massachusetts Institute of Technology, Cambridge, MA 02139, USA}
\affiliation[b]{Institute for Particle Physics Phenomenology, Durham University, South Road, Durham DH1 3LE, United Kingdom}
\affiliation[c]{IPhT, CEA Saclay, CNRS URA 2306, F-91191 Gif-sur-Yvette, France}

\emailAdd{larkoski@mit.edu}
\emailAdd{simone.marzani@durham.ac.uk}
\emailAdd{gregory.soyez@cea.fr}
\emailAdd{jthaler@mit.edu}

\abstract{
We introduce a new jet substructure technique called ``soft drop declustering'', which recursively removes soft wide-angle radiation from a jet.  The soft drop algorithm depends on two parameters---a soft threshold $\zcut$ and an angular exponent $\beta$---with the $\beta = 0$ limit corresponding roughly to the (modified) mass drop procedure.  To gain an analytic understanding of soft drop and highlight the $\beta$ dependence, we perform resummed calculations for three observables on soft-dropped jets:  the energy correlation functions, the groomed jet radius, and the energy loss due to soft drop.  The $\beta = 0$ limit of the energy loss is particularly interesting, since it is not only ``Sudakov safe'' but also largely insensitive to the value of the strong coupling constant.  While our calculations are strictly accurate only to modified leading-logarithmic order, we also include a discussion of higher-order effects such as multiple emissions and (the absence of) non-global logarithms.  We compare our analytic results to parton shower simulations and find good agreement, and we also estimate the impact of non-perturbative effects such as hadronization and the underlying event.  Finally, we demonstrate how soft drop can be used for tagging boosted $W$ bosons, and we speculate on the potential advantages of using soft drop for pileup mitigation.
}

\begin{document} 
\maketitle

\section{Introduction}

The study of jet substructure has significantly matured over the past five years \cite{boost2010,boost2011,boost2012}, with numerous techniques proposed to tag boosted objects  \cite{Seymour:1993mx,Butterworth:2002tt,BDRS,Almeida:2008yp,Almeida:2008tp,Kaplan:2008ie,Brooijmans:1077731,Thaler:2008ju,CMS:2009lxa,CMS:2009fxa,Rappoccio:1358770,ATL-PHYS-PUB-2009-081,Kribs:2009yh,Kribs:2010hp,Chen:2010wk,Hackstein:2010wk,Falkowski:2010hi,Katz:2010mr,Cui:2010km,Kim:2010uj,Thaler:2010tr,Gallicchio:2010dq,Gallicchio:2010sw,ATL-PHYS-PUB-2010-008,Plehn:2009rk,Plehn:2010st,Almeida:2010pa,Thaler:2011gf,Jankowiak:2011qa,Hook:2011cq,Soper:2011cr,Almeida:2011aa,Ellis:2012sn,Soper:2012pb,Backovic:2012jj,Cohen:2012yc,Curtin:2012rm,Hedri:2013pvl,Backovic:2013bga,Gouzevitch:2013qca,Larkoski:2013eya,Chien:2013kca,Kahawala:2013sba}, distinguish quark from gluon jets \cite{Gallicchio:2011xq,Gallicchio:2012ez,Krohn:2012fg,Chatrchyan:2012sn,Pandolfi:1480598,Larkoski:2013eya}, and mitigate the effects of jet contamination \cite{Cacciari:2007fd,BDRS,trimming,Ellis:2009su,Ellis:2009me,Alon:2011xb,Soyez:2012hv,semi-classical,taggersRES,taggersNLO,Krohn:2013lba}.  Many of these techniques have found successful applications in jet studies at the Large Hadron Collider (LHC) \cite{Miller:2011qg,Aad:2012raa,ATLAS:2012dp,Aad:2012meb,ATLAS-CONF-2011-073,ATLAS-CONF-2011-053,ATLAS:2012am,ATLAS:2012jla,ATLAS-CONF-2012-065,ATLAS-CONF-2012-066,Aad:2013fba,Aad:2013gja,TheATLAScollaboration:2013pia,TheATLAScollaboration:2013qia,TheATLAScollaboration:2013ria,TheATLAScollaboration:2013sia,TheATLAScollaboration:2013tia,ATLAS:2012xna,CMS-PAS-JME-10-013,CMS:2011bqa,CMS-PAS-QCD-10-041,Chatrchyan:2012tt,CMS:2013kfa,CMS:2013wea,CMS:2013vea,CMS:2013uea,Chatrchyan:2012mec,Chatrchyan:2012sn,cms}, and  jet substructure is likely to become even more relevant with the anticipated increase in energy and luminosity for Run II of the LHC.

In addition to these phenomenological and experimental studies of jet substructure, there is a growing catalog of first-principles calculations using perturbative QCD (pQCD).  These include more traditional jet mass and jet shape distributions \cite{Li:2012bw,Ellis:2010rwa,BDKM,DKMS,Chien:2012ur,Jouttenus:2013hs} as well as more sophisticated substructure techniques \cite{Rubin:2010fc,Walsh:2011fz,Feige:2012vc,Field:2012rw,Larkoski:2012eh,Larkoski:2013eya,Gerwick:2012fw,taggersRES, taggersNLO,Larkoski:2014uqa,Larkoski:2014tva}.   Recently, \Refs{taggersRES, taggersNLO} considered the analytic behavior of three of the most commonly used jet tagging/grooming methods---trimming \cite{trimming}, pruning \cite{Ellis:2009su,Ellis:2009me}, and mass drop tagging \cite{BDRS}.  Focusing on groomed jet mass distributions, this study showed how their qualitative and quantitative features could be understood with the help of logarithmic resummation.  Armed with this analytic understanding of jet substructure, the authors of \Ref{taggersRES} developed the modified mass drop tagger (mMDT) which exhibits some surprising features in the resulting groomed jet mass distribution, including the absence of Sudakov double logarithms, the absence of non-global logarithms \cite{non-global}, and a high degree of insensitivity to non-perturbative effects.

In this paper, we introduce a new tagging/grooming method called ``soft drop declustering'', with the aim of generalizing (and in some sense simplifying) the mMDT procedure.  Like any grooming method, soft drop declustering removes wide-angle soft radiation from a jet in order to mitigate the effects of contamination from initial state radiation (ISR), underlying event (UE), and multiple hadron scattering (pileup).  Given a jet of radius $R_0$ with only two constituents, the soft drop procedure removes the softer constituent unless
\begin{equation}
\label{eq:mainrelation}
\text{Soft Drop Condition:} \quad \frac{\min(p_{T1},p_{T2})}{p_{T1}+p_{T2}} > \zcut \left(\frac{\Delta R_{12}}{R_0} \right)^\beta,
\end{equation}
where $p_{Ti}$ are the transverse momenta of the constituents with respect to the beam, $\Delta R_{12}$ is their distance in the rapidity-azimuth plane, $\zcut$ is the soft drop threshold, and $\beta$ is an angular exponent.  By construction, \Eq{eq:mainrelation} fails for wide-angle soft radiation.  The degree of jet grooming is controlled by $\zcut$ and $\beta$, with $\beta \to \infty$ returning back an ungroomed jet.  As we explain in \Sec{sec:softdrop}, this procedure can be extended to jets with more than two constituents with the help of recursive pairwise declustering.\footnote{\label{footnote:semiclassical}The soft drop procedure takes some inspiration from the ``semi-classical jet algorithm''~\cite{semi-classical}, where a variant of \Eq{eq:mainrelation} with $\zcut = 1/2$ and $\beta = 3/2$ is tested at each stage of recursive clustering (unlike declustering considered here).}

Following the spirit of \Ref{taggersRES}, the goal of this paper is to understand the analytic behavior of the soft drop procedure, particularly as the angular exponent $\beta$ is varied.  There are two different regimes of interest.  For $\beta > 0$, soft drop declustering removes soft radiation from a jet while still maintaining a fraction (controlled by $\beta$) of the soft-collinear radiation.  One of the consequences is that the soft drop procedure gives infrared/collinear (IRC) safe results even on a jet with just one constituent.  In this regime, soft drop acts like a ``groomer'', meaning that it changes the constituents of a jet without affecting the overall  jet production cross section.  For $\beta < 0$, soft drop declustering can remove both soft and collinear radiation.  For a jet to pass the soft drop procedure, it must have at least two constituents satisfying \Eq{eq:mainrelation}.  Thus, in this regime, soft drop acts like a ``tagger'', since it vetoes jets that do not have two well-separated hard prongs.  Roughly speaking, the boundary $\beta = 0$ corresponds to mMDT, which acts like a tagger at any fixed-order in an $\alpha_s$ expansion, but can be thought of as a ``Sudakov safe'' \cite{Larkoski:2013paa} groomer when all orders in $\alpha_s$ are considered.

To demonstrate the behavior of the soft drop procedure, we will present three calculations performed on soft-dropped jets.
\begin{itemize}
\item \textit{Energy correlation functions}.  The generalized energy correlation functions (ECF) were introduced in~\Ref{Larkoski:2013eya}, where $\ECF{N}{\alpha}$ corresponds to an $N$-point correlation function with angular exponent $\alpha$.  In this paper, we will focus on the 2-point correlator through the combination $\C{1}{\alpha} \equiv  \ECF{2}{\alpha} /  \ECF{1}{\alpha}^2$ (see also \Refs{Banfi:2004yd,Jankowiak:2011qa}).  For a jet with two constituents,
\begin{equation}
\label{eq:C1simpdef}
\C{1}{\alpha} \simeq \frac{p_{T1} \, p_{T2}}{(p_{T1} + p_{T2})^2}  \left(\frac{\Delta R_{12}}{R_0} \right)^\alpha,
\end{equation}
where we have added an extra $R_0$ normalization factor for later convenience.  The value $\alpha = 2$ is related to jet thrust/mass \cite{Berger:2003iw,Almeida:2008yp,Ellis:2010rwa}, $\alpha = 1$ is related to jet broadening/girth/width \cite{Gallicchio:2010dq,Gallicchio:2011xq}, and arbitrary $\alpha > 0$ is related to the recoil-free angularities \cite{Larkoski:2014uqa}.  In \Sec{sec:ang}, we calculate $\C{1}{\alpha}$ in the modified leading logarithmic (MLL) approximation, which accounts for all terms $\as^n L^{2n-q}$ with $q=0,1$ and $L \equiv \log (1/\ea)$ in the expansion of the $\C{1}{\alpha}$ cumulative distribution.
We will also compute higher-order effects due to multiple emissions and we will find an interesting interplay between the ECF exponent $\alpha$ and the soft drop exponent $\beta$, especially as relates to non-global logarithms.
\item \textit{Groomed jet radius}.  The soft drop declustering procedure terminates when \Eq{eq:mainrelation} is satisfied, and the corresponding $\Delta R_{12}$ gives the effective radius $R_g$ of the groomed jet.  Roughly speaking, the active jet area \cite{jet-area} is $ \simeq \pi R_g^2$.  In \Sec{sec:pileup}, we calculate the $R_g$ distribution to MLL accuracy to gain an understanding of how the soft drop procedure might perform in a pileup environment.
\item \textit{Jet energy drop}.  Strictly speaking, the groomed jet energy distribution after mMDT (i.e.\ $\beta = 0$) is not IRC safe. One of the motivations for introducing the generalized soft drop procedure with $\beta > 0$ is to have a method (in the same spirit of trimming \cite{trimming}) that gives IRC safe distributions for any (otherwise) IRC safe observable measured on groomed jets.  In \Sec{sec:energy}, we calculate the fractional drop in the jet energy after the soft drop procedure to MLL accuracy, including higher-order corrections due to multiple emissions. Intriguingly, we will find that the $\beta \to 0$ limit is ``Sudakov safe'' \cite{Larkoski:2013paa}, and the resulting jet energy drop spectrum is \emph{independent} of  $\alpha_s$ in the fixed coupling approximation.
\end{itemize}
While the focus of this paper is on the analytic properties of the soft drop procedure, we will cross check our results using parton shower Monte Carlo simulations.  In addition to these analytic studies, we will perform a Monte Carlo study of non-perturbative corrections (hadronization and UE) in \Sec{sec:NP}, and estimate the tagging performance of soft drop for boosted $W$ bosons in \Sec{sec:tagging}. We present our conclusions in \Sec{sec:conclude}.

\section{Soft Drop Declustering}
\label{sec:softdrop}

\subsection{Definition}
\label{sec:def}

The starting point for soft drop declustering is a jet with characteristic radius $R_0$.  For definiteness, we will always consider jets defined with the anti-$k_t$ algorithm~\cite{anti-kt}, but other jet algorithms would work equally well.  We then recluster the jet constituents using the Cambridge-Aachen (C/A) algorithm~\cite{Dokshitzer:1997in, Wobisch:1998wt} to form a pairwise clustering tree with an angular-ordered structure.

The soft drop declustering procedure depends on two parameters, a soft threshold $\zcut$ and an angular exponent $\beta$, and is implemented as follows:
\begin{enumerate} 
\item Break the jet $j$ into two subjets by undoing the last stage of C/A clustering.  Label the resulting two subjets as $j_1$ and $j_2$.
\item \label{step:grooming} If the subjets pass the soft drop
  condition \Big($\frac{\min(p_{T1},p_{T2})}{p_{T1}+p_{T2}}>\zcut
  \left(\frac{\Delta R_{12}}{R_0} \right)^\beta$, see
  \Eq{eq:mainrelation}\Big) then deem $j$ to be the final soft-drop
  jet.  (Optionally, one could also impose the mass-drop condition $\max(m_1,m_2)<\mu \,m$ as in \Ref{BDRS}, but we will not use that here.)
\item Otherwise, redefine $j$ to be equal to subjet with larger $p_T$ and iterate the procedure.
\item If $j$ is a singleton and can no longer be declustered, then one can either remove $j$ from consideration (``tagging mode'') or leave $j$ as the final soft-drop jet (``grooming mode'').
\end{enumerate}
By building a C/A tree, we can apply the pairwise soft drop condition from \Eq{eq:mainrelation} to a jet with more than two constituents.  Tagging mode is only IRC safe for $\beta \le 0$ whereas grooming mode is only IRC safe for $\beta > 0$.  In this paper, we will typically consider $\zcut \simeq 0.1$ but we will explore a wide range of $\beta$ values.\footnote{Throughout this paper, we will assume that $\Delta R_{12}<{R_0}$ at every stage of the declustering, such that the algorithm returns the whole jet in the $\beta \to \infty$ limit.  In practice, it is possible for a jet of characteristic radius $R_0$ to have $\Delta R_{12}> {R_0}$ when reclustered with C/A, and in that case we simply apply step~\ref{step:grooming} without change, such that wide angle emissions can still be vetoed even in the $\beta \to \infty$ limit.}

The above algorithm can be thought of as a generalization of the (modified) mass-drop tagger (mMDT)~\cite{BDRS,taggersRES}, with $\beta = 0$ roughly corresponding to mMDT itself.  There are, however, a few important differences.  First, soft drop declustering does not require a mass drop condition (or equivalently, the mass drop parameter $\mu$ is set to unity).  As shown in \Ref{taggersRES}, the mass drop condition is largely irrelevant for understanding the analytic behavior of mMDT on quark/gluon jets, so we have decided not to include it in the definition here.  Second, we note that the $\beta = 0$ limit corresponds to a mMDT variant where step~\ref{step:grooming} is implemented directly on the transverse momentum fractions of subjets, rather than indirectly through a ratio of a $k_t$-distance to a mass~\cite{taggersRES}.  Of course, the two give the same behavior in the small $\min(p_{T1},p_{T2})/(p_{T1}+p_{T2})$ limit, but \Eq{eq:mainrelation} makes it obvious that the soft drop condition drops soft radiation (true to its name).  Finally and most importantly, for $\beta \not= 0$, the soft drop condition involves a relation between energies and angular distances, rather than just energies as is the case for $\beta = 0$.  It is this additional angular dependence (exploited by the exponent $\beta$) that we wish to highlight in this paper.

As mentioned in footnote \ref{footnote:semiclassical}, the soft drop condition takes some inspiration from the ``semi-classical jet algorithm''~\cite{semi-classical}.  The semi-classical algorithm is a pairwise clustering algorithm that only allows mergings which satisfy
\begin{equation}
\text{Semi-classical Condition:} \quad  \frac{\min(m_{T1},m_{T2})}{m_{T1} + m_{T2}} > \frac{1}{2} \left(\frac{\Delta R_{12}}{R_0}  \right)^{3/2},
\end{equation}
where $m_{Ti} = \sqrt{m_i^2 + p_{Ti}^2}$.  Apart from the change of $p_{Ti} \to m_{Ti}$, the semi-classical condition looks like the soft drop condition with $\beta = 3/2$ and $\zcut = 1/2$, but there is an important difference.  For semi-classical jets, one is recursively clustering a jet using a novel measure.  For soft-drop jets, one is taking an existing jet defined with a traditional algorithm and using soft drop declustering to groom away soft wide-angle emissions.  Of course, the distinction between clustering and declustering is irrelevant for a jet with only two constituents, but it is very important for our analytic calculations which only apply to declustering of a C/A tree.\footnote{In principle, it is possible to use any of the generalized $k_t$ algorithms \cite{Catani:1993hr,Ellis:1993tq} to perform the soft drop declustering. The choice of C/A is motivated by the approximate angular ordering of emissions in the parton shower.}

\begin{figure}
\begin{center}
\includegraphics[scale=0.65]{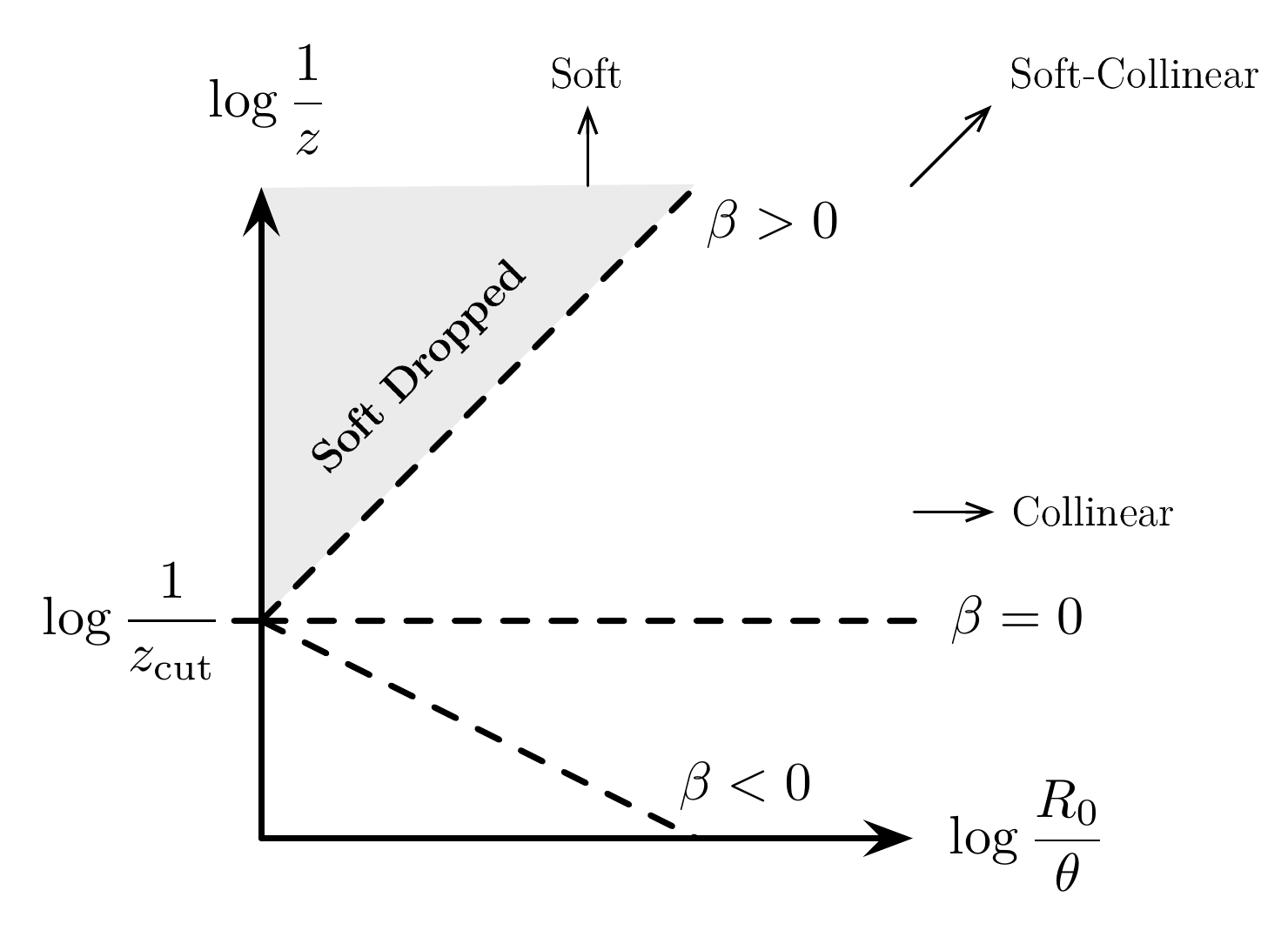}
\end{center}
\caption{Phase space for emissions on the $(\log \frac{1}{z},\log \frac{R_0}{\theta})$ plane.  In the strongly-ordered limit, emissions above the dashed line (\Eq{eq:vetoline}) are vetoed by the soft drop condition.  For $\beta > 0$, soft emissions are vetoed while much of the soft-collinear region is maintained.  For $\beta = 0$ (mMDT), both soft and soft-collinear emissions are vetoed.  For $\beta < 0$, all (two-prong) singularities are regulated by the soft drop procedure.}
\label{fig:groomedregions}
\end{figure}

\subsection{Dependence on $\beta$}
\label{sec:betachange}

Before studying the analytic behavior of soft-drop distributions in detail, it is worth making a few general comments about the expected $\beta$ dependence.   For simplicity of discussion, we will work with central jets (i.e.\ rapidity $y = 0$) with small radius ($R_0 \ll 1$).  This way, we can freely exchange transverse momentum $p_T$ for energy $E$, as well as rapidity-azimuth distance $R$ for opening angle $\theta$.  All of the results of this paper extend to non-zero rapidity as well, up to power corrections in the jet radius, which we neglect.

In \Fig{fig:groomedregions}, we show the phase space for a single gluon emission from an eikonal hard quark/gluon on the $(\log \frac{1}{z},\log \frac{R_0}{\theta})$ plane, where $0 \le z \le 1$ is the energy fraction and $0 \le \theta \le R_0$ is the angle of the emission. We have labeled three modes in the phase space: soft, soft-collinear, and collinear.  For this paper, we define the various modes in terms of their $z$ and $\theta$ behavior: 
\begin{align*}
\text{soft modes:} &\quad z\to 0,\ \theta = \text{constant}, \\
\text{soft-collinear modes:}  &\quad z\to 0,\ \theta \to 0, \\
\text{collinear modes:}  &\quad z=\text{constant},\ \theta \to 0. 
\end{align*}
No relative scaling is assumed between energy fraction $z$ and splitting angle $\theta$ for soft-collinear modes. 
In these logarithmic coordinates, the emission probability is flat in the soft-collinear limit.  In the soft limit, the soft drop criteria reduces to
\begin{equation}
\label{eq:vetoline}
z > \zcut \left(\frac{\theta}{R_0}\right)^\beta \quad \Rightarrow \quad \log \frac{1}{z} < \log \frac{1}{\zcut} + \beta \log \frac{R_0}{\theta}.
\end{equation}
Thus, vetoed emissions lie above a straight line of slope $\beta$ on the $(\log \frac{1}{z},\log \frac{R_0}{\theta})$ plane, as shown in \Fig{fig:groomedregions}.

For $\beta > 0$, collinear radiation always satisfies the soft drop condition, so a soft-drop jet still contains all of its collinear radiation.  The amount of soft-collinear radiation that satisfies the soft drop condition depends on the relative scaling of the energy fraction $z$ to the angle $\theta$.  As $\beta\to 0$, more of the soft-collinear radiation of the jet is removed, and in the $\beta=0$ (mMDT) limit, all soft-collinear radiation is removed.  Therefore, we expect that the coefficient of the double logarithms of observables like groomed jet mass (and $\C{1}{\alpha}$) will be proportional to $\beta$, when $\beta$ is small.
Similarly, because the soft drop procedure does not change the structure of collinear emissions, observables like the groomed jet energy are IRC safe. Note that running $\beta > 0$ soft drop in tagging mode is not IRC safe, since a jet would (would not) be tagged if it contained two (one) collinear particles. 

In the strict $\beta = 0$ or mMDT limit, collinear radiation is only maintained if $z > \zcut$.  Because soft-collinear radiation is vetoed, the resulting jet mass (and $\C{1}{\alpha}$) distributions will only exhibit single logarithms, as emphasized in \Refs{taggersRES,taggersNLO}. Because the structure of collinear emissions is modified, observables like groomed jet energy are only IRC safe if soft drop is used in tagging mode, since that forces the jet to have a hard two-prong structure, which regulates the collinear singularity.  We will see in \Sec{sec:energy}, however, that $\beta = 0$ grooming mode is still ``Sudakov safe'' \cite{Larkoski:2013paa}.

Finally, for $\beta < 0$, there are no logarithmic structures for observables like groomed jet mass at arbitrarily low values of the observable.  Effectively, soft drop with negative $\beta$ acts like a cut which enforces
 $\C{1}{\alpha} > \zcut^{\alpha/|\beta|}$,
and this cut regulates the soft-collinear singularities.  For example, $\beta = -1$ roughly corresponds to a cut on the relative transverse momentum of the two prongs under scrutiny.  Like for $\beta = 0$, $\beta < 0$ is only IRC safe in tagging mode.

\section{Energy Correlation Functions after Soft Drop} \label{sec:ang}

Generalized energy correlation functions $\ECF{N}{\alpha}$ and their double ratios $\C{N-1}{\alpha}$ were introduced in~\Ref{Larkoski:2013eya} (see also \Refs{Banfi:2004yd,Jankowiak:2011qa} for $N = 2$). In this paper, we only consider the double ratio for $N=2$ (hereafter referred to as simply the energy correlation functions):
\begin{equation}\label{C1b-def}
\C{1}{\alpha}= \frac{\ECF{2}{\alpha} \ECF{0}{\alpha}}{\ECF{1}{\alpha}^2},
\end{equation}
where
\begin{eqnarray} \label{ECF-defs}
\ECF{0}{\alpha}&=& 1, \nonumber \\
\ECF{1}{\alpha}&=& \sum_{i \in  \text{jet}} p_{Ti}, \nonumber \\
\ECF{2}{\alpha}&=& \sum_{i <j \, \in \text{jet}} p_{Ti} \, p_{Tj} \left( \frac{\Delta R_{ij}}{R_0}\right)^\alpha.
\end{eqnarray}
In this study, we will measure $\C{1}{\alpha}$ on jets which have been groomed according to the soft-drop declustering described above.  We will work to lowest non-trivial order in $z_{\rm cut}$, such that we can ignore the effect of grooming on $\ECF{1}{\alpha}$.  As stated above, we will focus on central jets ($y = 0$) and assume $R_0 \ll 1$.  In those limits,
\begin{equation}
\C{1}{\alpha} \simeq \sum_{i < j} z_i z_j \left( \frac{\theta_{ij}}{R_0}\right)^\alpha,
\end{equation}
where $z_i \simeq E_i / E_{\text{jet}}$ is the energy fraction carried by particle $i$, and $\theta_{ij}$ is the opening angle between particles $i$ and $j$.  Up to power-suppressed effects in $R_0$, the results of this paper can be extended to non-zero rapidity ($y \not= 0$) by simply replacing $\theta_{ij}$ with the rapidity-azimuth distance $R_{ij}$ and the energy fraction $z_i$ with the momentum fraction $p_{Ti}/ p_{T\text{jet}}$.

\subsection{Leading-Order Calculation}
\label{sec:ang-lo}

We start our analysis with a relatively simple calculation, by computing the leading order (LO) contribution to the $\ea$ distribution in the collinear limit.  This limit is appropriate for the small $R_0$ assumption considered throughout this paper. 

At LO, the jet consists of only two partons at an angular distance $\Delta R_{12}\simeq \theta$, which carry fractions $z$ and $(1-z)$ of the jet's energy.  To have a non-zero contribution to $\ea$, both partons must pass the soft-drop condition.  In the collinear limit, the groomed $\ea$ distribution is
\begin{multline}\label{ang-LO-start}
\frac{1}{\sigma}\frac{d \sigma^\text{LO}}{d \ea}=\frac{ \as}{\pi} \int_0^{R_0} \frac{d \theta}{\theta} \int_0^1 d z\, p_i(z) \, 
\Theta\left(z-\zcut \left(\frac{\theta}{R_0} \right)^\beta \right)\Theta\left(1-z-\zcut \left(\frac{\theta}{R_0} \right)^\beta \right)  \\ ~\times \delta \left(\ea - z(1-z) \left(\frac{\theta}{R_0} \right)^\alpha \right),
\end{multline}
where $p_i(z)$ is the appropriate splitting function for a quark-initiated jet ($i=q$) or a gluon-initiated jet ($i=g$), as defined in \Eq{eq:reduced-splitting}.  The two theta functions impose the soft drop condition, and the delta function implements the $\ea$ measurement.

Because we work in the limit where $\ea\ll\zcut\ll1$, we can ignore terms suppressed by powers of $\zcut$ (but we do not need to resum logarithms of $\zcut$); this implies that we can ignore the second theta function in \Eq{ang-LO-start}.  Only focusing on the logarithmically-enhanced contributions, we can also drop the factor of ($1-z$) in the delta function.  These simplifications lead to
\begin{equation}
\label{ang-LO-mid}
\frac{1}{\sigma}\frac{d \sigma^\text{LO}}{d \ea} \simeq \frac{\as}{\pi} \int_0^{R_0} \frac{d \theta}{\theta} \int_0^1 d z\, p_i(z) \, 
\Theta\left(z-\zcut \left(\frac{\theta}{R_0} \right)^\beta \right) \, \delta \left(\ea - z \left(\frac{\theta}{R_0} \right)^\alpha \right).
\end{equation}
For $\beta \ge 0$, the evaluation of the two integrals is straightforward:
\begin{equation}
\label{ang-LO-res}
\beta \ge 0: \quad \frac{\ea}{\sigma}\frac{d \sigma^\text{LO}}{d \ea} \simeq \frac{\as C_i}{\pi}\frac{2}{\alpha} \times
\begin{cases}
\log \frac{1}{\ea}+ B_i , & \ea > \zcut, \\
\frac{\beta}{\alpha+\beta} \log \frac{1}{\ea}+ \frac{\alpha}{\alpha+\beta}\log \frac{1}{\zcut} +B_i, & \ea < \zcut,
\end{cases}
\end{equation}
up to terms that are power-suppressed in $\ea$ or $\zcut$.
Here, $C_i$ is the overall color factor for the jet ($C_q = C_F = 4/3$ for quarks and $C_g = C_A = 3$ for gluons) and $B_i$ originates from hard-collinear emissions ($B_q = -3/4$ for quarks and $B_g = -\frac{11}{12}+\frac{n_f}{6C_A} $ for gluons, where $n_f$ is the number of active quark flavors).  For $\beta < 0$, there is an additional restriction which imposes a minimum allowed value for the observable
\begin{equation}
\label{eq:negbetaminvalue}
\beta < 0:  \quad \text{Same as \Eq{ang-LO-res} with additional cut } \ea> \zcut^{\alpha/ |\beta|}.
\end{equation}

As often happens for grooming and tagging algorithms~\cite{taggersRES, taggersNLO}, the $\ea$ distribution exhibits a transition point at $\ea=\zcut$.  Unlike trimming and pruning, though, soft-drop energy correlation functions do not exhibit further (perturbative) transition points at lower values of the observable.   For $\ea> \zcut$, soft drop is not active and we recover the ungroomed result.  For $\ea < \zcut$, soft drop is active and jets that fail the soft drop condition are either removed from consideration (tagging mode) or assigned $\ea = 0$ (grooming mode).  Note that for $\beta > 0$, the logarithmic structure of \Eq{ang-LO-res} is of the same order on both sides of the transition point, so the overall cumulative distribution exhibits Sudakov double logarithms.  The effect of the soft drop procedure is to reduce the coefficient of the double logarithm by a factor of $\beta/(\alpha + \beta)$.

It is instructive to take different limits of the result in \Eq{ang-LO-res}.  Consider the $\beta\to \infty$ limit at fixed $\alpha$ and $\zcut$.  This limit should correspond to no grooming, and indeed, in this limit, we recover the expected LO result for the energy correlation function of the ungroomed jet.  Now consider the case $\beta = 0$, which should correspond to the mMDT limit.  This limit kills the logarithmic contribution for $\ea < \zcut$, which results in a cumulative distribution that only has single logarithms in $\ea$. This result is the generalization to $\ea$ of the fact that the mMDT jet mass distribution (here $\alpha=2$) is only single logarithmic~\cite{taggersRES, taggersNLO}.

\subsection{Modified Leading Logarithmic Approximation}
\label{sec:MLL}

Because of the potentially large logarithms $L \equiv \log (1/\ea)$ in \Eq{ang-LO-res}, we need to perform some kind of resummation in order to get realistic predictions for the $\ea$ distribution.  Here, we investigate a simple approximation to the all-order $\ea$ distribution by working to modified leading logarithmic (MLL) accuracy, i.e.\ we aim to capture the terms $\as^n L^{2n-q}$ with $q=0,1$ in the expansion of the cumulative distribution $\Sigma(\ea)$, which gives the probability for the observable to be less than a given value $\ea$.

\begin{figure}
\begin{center}
\includegraphics[scale=0.55]{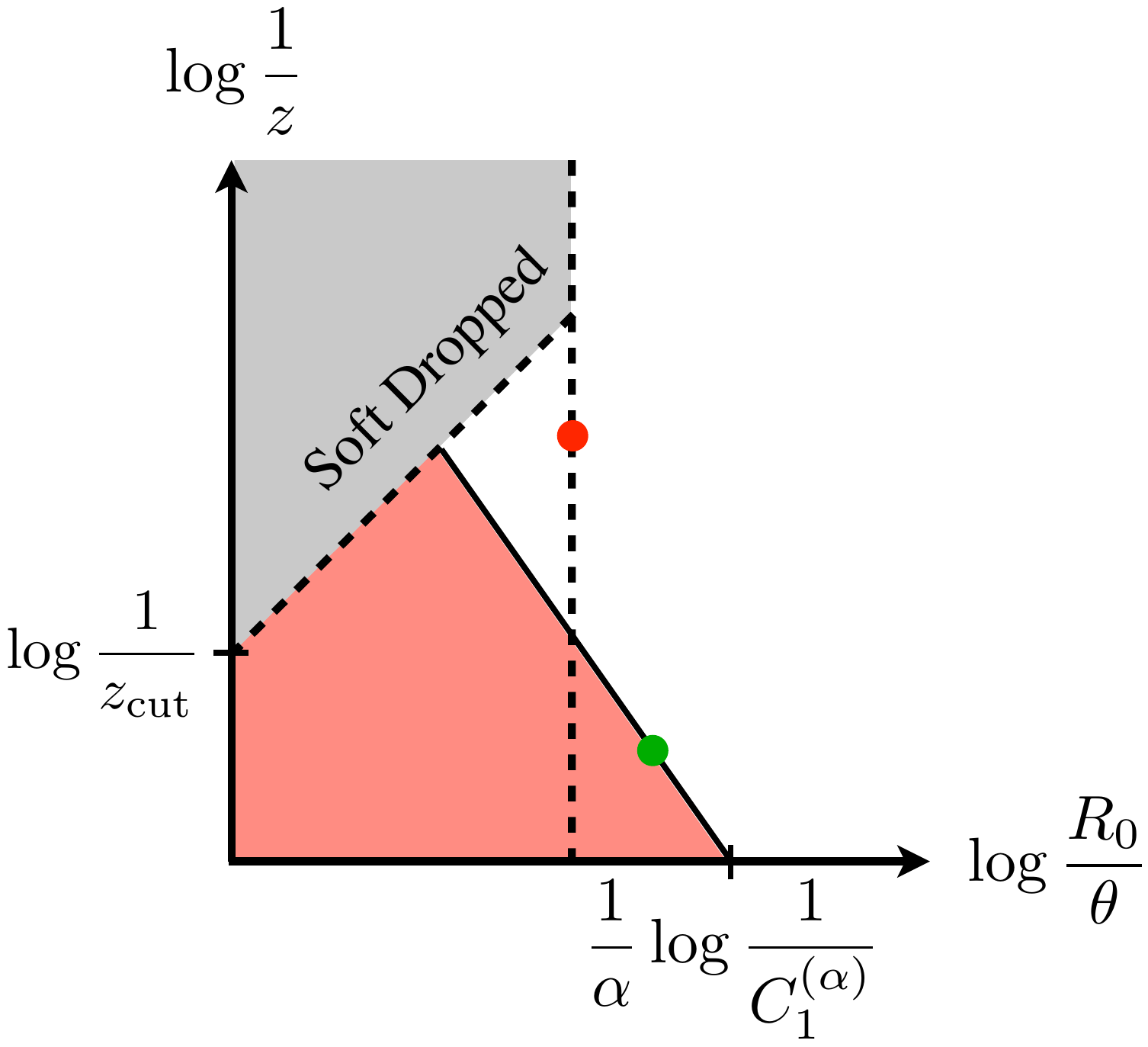}
\end{center}
\caption{Phase space for emissions relevant for $\ea$ in the $(\log \frac{1}{z},\log \frac{R_0}{\theta})$ plane.  The soft dropped region is gray and the first emission satisfying the soft drop criteria is illustrated by the red dot.  The leading emission for $\ea$ is illustrated by the  green dot with the forbidden emission region (the Sudakov exponent) shaded in pink.
}
\label{fig:groomedregions-C1}
\end{figure}

To MLL order, we need to consider the independent emission of any number of soft or collinear gluons within a jet, with the scale of the (one-loop) coupling chosen at the relative transverse momentum scale $\kappa$ of the splitting.  In the collinear approximation used throughout this paper, we have $\kappa = z \, \theta \, p_{T \text{jet}}$ for jets at arbitrary rapidity.

Virtual emissions are associated with $\ea = 0$ and therefore always contribute to $\Sigma(\ea)$.  A real emission contributes to $\Sigma(\ea)$ either if it has been groomed away or if it lies at an angle smaller than the first emission that passes the soft drop condition.  This is illustrated in \Fig{fig:groomedregions-C1}.  Note that the dominant emission contributing to the value of $\ea$ must always lie at an angle less than or equal to the angle of the first emission that passes the soft drop condition, so we do not need to consider the correlation of the groomed jet radius and the value of $\ea$.

The details of the MLL calculation and explicit results are presented in \App{app:angdist}.  After one explicitly does the sum over all included real emissions, the cumulative distribution can be interpreted in terms of the phase space for vetoed real emissions.  This gives the Sudakov exponent
\begin{equation}\label{all-order-ang-end}
\Sigma(\ea)=\exp \left[- \frac{1}{\alpha}\int_{\ea}^1 \frac{d c}{c} 
\int_{\max \left(c,\, {\zcut}^\frac{\alpha}{\alpha+\beta} {c}^\frac{\beta}{\alpha+\beta} \right) }^1 d z \, p_i(z)
\frac{\as\left(\kappa  \right)}{\pi} 
\right] \ ,
\end{equation}
where we have introduced the convenient change of variables $c= z \left( \frac{\theta}{R_0}\right)^\alpha$.  The integral in the exponent corresponds to real emissions that are not removed by the soft drop procedure, but would give a too large contribution to $\ea$ (corresponding to the pink shaded region in \Fig{fig:groomedregions-C1}).  As expected, the $\beta\to \infty$ limit corresponds to the ungroomed result, and the $\beta = 0$ (mMDT) limit matches the jet mass ($\alpha = 2$) distribution in \Ref{taggersRES}.

To better understand the logarithmic structure of the soft-dropped energy correlation functions, it is instructive to perform the integrals in \Eq{all-order-ang-end} in a fixed coupling approximation. For $\beta \ge 0$, neglecting power-suppressed terms, we obtain \newpage
\begin{multline}\label{all-order-ang-fc}
\Sigma(\ea)
 \overset{\rm f.c.}{\wideeq{0.6cm}}
\exp \Bigg\{- \frac{\as C_i}{\pi}\frac{2}{\alpha}
\Bigg[ \left( \frac{1}{2}\log^2 \frac{1}{\ea}+ B_i \log\frac{1}{\ea} \right) \Theta\left(1- \ea \right)  \Theta\left(\ea - \zcut \right)  \\
+
\left( \frac{\beta}{2(\alpha+\beta)} \log^2 \frac{1}{\ea}+ \frac{\alpha}{\alpha+\beta}\log \frac{1}{\zcut} \log\frac{1}{\ea}-\frac{\alpha}{2(\alpha+\beta)} \log^2 \frac{1}{\zcut} +B_i \log\frac{1}{\ea}\right) \\ \times \Theta\left(\zcut-\ea \right)  
\Bigg] \Bigg\}  \,,
\end{multline}
which is the exponential of the cumulative distribution at LO (i.e.\ the integral of \Eq{ang-LO-res}).  For $\beta < 0$, we find an expression analogous to \Eq{all-order-ang-fc}, but with a lower bound which enforces $\C{1}{\alpha} > \zcut^{\alpha/|\beta|}$, thus regulating the soft-collinear behavior.  The limiting values of \Eq{all-order-ang-fc} behave as expected.  For $\beta \to \infty$, the regions above and below $\zcut$ give identical results, so they can be combined to return the ungroomed distribution. For $\beta=0$, the coefficient of the double logarithm in the region $\ea<\zcut$ vanishes and we obtain the expected mMDT single logarithmic result. 

\subsection{Multiple Emissions}
\label{sec:pme}

Multiple gluon emissions within a jet can affect the value of $\ea$. While this effect is strictly speaking beyond MLL accuracy, it is an important component of a full NLL calculation, so it is worth considering how they might affect the $\ea$ distribution.  For multiple emissions, we need to determine what region of phase space can have several emissions that contribute to the measured value of the observable.  To logarithmic accuracy, these emissions must give comparable contributions to the final measured value of the observable.

For the case of the energy correlation function $\ea$, the region of phase space where multiple emissions contribute can be seen in \Fig{fig:groomedregions-C1}.  For the green emission that sets the value of $\ea$, multiple emissions that contribute logarithmically must lie near the diagonal line defining a fixed value for $\ea$.  Everywhere along this diagonal line satisfies the soft drop groomer, and therefore all emissions that contribute to the value of $\ea$ pass the soft drop phase space requirements.\footnote{One might worry that if the emission that sets the value of $\ea$ lies near the boundary between the soft-drop groomed region and soft-drop kept region, then emissions that contribute to the observable may not satisfy the soft-drop requirement on their own.  While this is true, the contributions from such emissions are subleading to the accuracy to which we work and can therefore be ignored.}  Also, because C/A clustering enforces angular ordering, these multiple emissions must lie at angles smaller than the first emission that passes the soft drop requirement.  Therefore, accounting for multiple emissions requires including an arbitrary number of emissions that contribute to $\ea$ and pass the soft drop requirement.  

To single logarithmic accuracy, the cumulative distribution of soft drop groomed $\ea$ can then be expressed as an explicit sum over uncorrelated emissions as 
\begin{align}
\Sigma(\ea)&= \sum_{n=1}^\infty \prod_{m=1}^n\left[ \int_0^{R_0} \frac{d\theta_m}{\theta_m}\int_0^1 dz_m \, p_i(z_m) \frac{\alpha_s(\kappa_m)}{\pi} \Theta\left(
z_m-\zcut\left(\frac{\theta_m}{R_0}\right)^\beta
\right)\Theta\left(
\theta_{i-1}-\theta_i
\right)\right]\nonumber \\
&\qquad\times\Theta\left(
\ea - \sum_{m=1}^n z_m \left(\frac{\theta_m}{R_0}\right)^\alpha
\right)e^{-\int_0^{R_0} \frac{d\theta}{\theta}\int_0^1 dz \, p_i(z) \frac{\alpha_s(\kappa)}{\pi} \Theta\left(
z-\zcut\left(\frac{\theta}{R_0}\right)^\beta
\right)} \ .
\end{align}
The requirement $\Theta\left(\theta_{i-1}-\theta_i\right)$ imposes angular ordering and the explicit exponential is the sum of virtual contributions.  The explicit sum can be evaluated by a Laplace transformation which yields
\begin{align}
\Sigma(\ea) = \int \frac{d\nu}{2\pi i \nu} \, e^{\nu \ea} e^{-R\left(\nu^{-1}\right) } \ ,
\end{align}
where the $\nu$ integral represents the inverse Laplace transform.  The function $R(\nu^{-1})$ is called the radiator and is
\begin{equation}
R\left(\nu^{-1}\right) = \int_0^{R_0} \frac{d\theta}{\theta}\int_0^1 dz\, p(z)\frac{\alpha_s(\kappa)}{\pi}\Theta\left(
z-\zcut\left(\frac{\theta}{R_0}\right)^\beta
\right)\left( 1-\exp\left[-\nu z\left( \frac{\theta}{R_0}  \right)^\alpha\right]  \right) \ .
\end{equation}
Because they are Laplace conjugates of one another, logarithmic accuracy in $\ea$ corresponds to the same logarithmic accuracy in $\nu$.  Therefore, for single logarithmic accuracy in $\ea$, we must compute the radiator to single logarithmic accuracy in $\nu$.  Expanding around $\nu^{-1}=\ea$, the inverse Laplace transform can be evaluated explicitly (see e.g.~\Ref{Catani:1992ua,Dokshitzer:1998kz,caesar}) and we find  
\begin{equation}\label{eq:multemis_c1}
\Sigma(\ea) = \frac{e^{-\gamma_E R'(\ea)}}{\Gamma\left(1+R'(\ea)\right)}e^{-R(\ea)} \ ,
\end{equation}
where 
\begin{equation} \label{eq:radiatorC1}
R(\ea) = \int_0^{R_0} \frac{d\theta}{\theta}\int_0^1 dz\, p(z)\frac{\alpha_s(\kappa)}{\pi}\Theta\left(
z-\zcut\left(\frac{\theta}{R_0}\right)^\beta
\right)\Theta\left(z\left( \frac{\theta}{R_0} \right)^\alpha-\ea \right)  \ ,
\end{equation}
$\gamma_E$ is the Euler-Mascheroni constant, $\Gamma$ is the gamma function, and
\begin{equation} \label{eq:R_der}
R'(\ea) = -\frac{\partial}{\partial \log \ea} R(\ea) \ .
\end{equation}
The prefactor in \Eq{eq:multemis_c1} containing $R'(\ea)$ captures the effect of multiple emissions on the distribution of $\ea$. We remind the reader that to single-logarithmic accuracy, we can neglect the hard-collinear contribution in the multiple-emission prefactor, i.e.\ we can take $B_i = 0$ in \Eq{eq:R_der}.

Multiple-emission contributions to the ungroomed $\ea$ distribution were considered in \Ref{Larkoski:2013eya}. The effect is non-negligible for the jet-mass like case ($\alpha=2$) and increases as $\alpha$ grows smaller. However, we expect these kind of contributions to be reduced by the soft-drop procedure, essentially because the coefficient of the soft-collinear terms, which give the single-logarithmic contribution to $R'$, is reduced by a factor $\order{\beta}$. We shall come back to this discussion in Sec.~\ref{sec:ang-MC}, when we compare the resummed calculation to a result obtained with a parton shower event generator.

The differential distribution for the observable $\ea$ with multiple emissions, i.e.\ the derivative of \Eq{eq:multemis_c1}, depends on the second derivative of the radiator function $R$. However, within our approximations, $R''$ is not continuous across $\ea=\zcut$ (see for instance \Eq{all-order-ang-fc}). Physically, this is a consequence of the fact that emissions that contribute similarly to the observable can occur on either side of the $\zcut$ transition point. As a result, the distribution with multiple emission exhibits a discontinuity at $\ea=\zcut$ because of terms which are beyond NLL accuracy in $\log \Sigma$.
In order to restore continuity, we can simply replace the logarithmic derivative with its discrete version:
\begin{equation} \label{finite-derivative}
R'(\ea) \to \frac{R(\ea \, e^{-\delta}) - R(\ea)}{\delta} \ .
\end{equation}
The specific choice of $\delta$ is irrelevant to single logarithmic accuracy, and we take $\delta = 1$ for definiteness.  One can think of the $\delta$-dependence as being one source of theoretical uncertainty.

\subsection{Non-Global Logarithms} \label{sec:ang-NGLs}

The jet-based $\ea$ is an example of a non-global observable~\cite{non-global}, meaning that it receives single-logarithmic contributions coming from an ensemble of gluons that are outside of the jet which then radiate soft gluons into the jet.
The resummation of non-global logarithms for the specific case of the mass of anti-$k_t$ jets ($\alpha=2$) was performed in~\Refs{BDKM, DKMS} in the large $N_C$ limit (for recent work at finite $N_C$ see \Ref{hatta}).  A key result of \Refs{taggersRES, taggersNLO} is that the mass distribution of an mMDT jet is free of non-global logarithms, since the mMDT eliminates all sensitivity to soft emissions.  Since non-global logarithms contribute only at the single-logarithmic level, they are formally beyond MLL accuracy.  That said, it is interesting to study the structure of non-global logarithms for soft-dropped $\ea$ as $\beta$ is varied, especially since we know non-global logarithms must vanish at $\beta = 0$.

Consider the lowest-order configuration that can produce a non-global logarithm, namely the correlated emission of two gluons where $k_1$ is outside the original anti-$k_t$ jet and a softer gluon $k_2$ is inside it.\footnote{Because the original jet is defined with the anti-$k_t$ algorithm, we are not sensitive to clustering logarithms first described in~\Ref{Banfi:2005gj}.}  To contribute to a non-global logarithm, $k_2$ has to pass the soft-drop condition, so the relevant phase space constraints are
\begin{equation}
\Theta^{\mathrm{NG}} \equiv \Theta \left (z_1-z_2 \right) \Theta \left( \theta_1-R_0 \right) \Theta \left(R_0-\theta_2 \right) \Theta \left(z_2-\zcut\left(\frac{\theta_2}{R_0} \right)^\beta  \right).
\end{equation}
To extract the non-global contribution, we have consider the $C_F C_A$ correlated emission term of the squared matrix element for two gluon emissions that satisfy the  $\Theta^{\mathrm{NG}}$ constraint:
\begin{equation} \label{ng-start}
\frac{1}{\sigma}\frac{d \sigma^\text{NG}}{d \ea}= 4 C_F C_A \left ( \frac{\alpha_s}{2 \pi} \right)^2 \int \frac{d z_1}{z_1} \frac{d z_2}{z_2} \int  \theta_2 d \theta_2 \int \theta_1 d \theta_1 \, \Omega_2 \, \Theta^{\mathrm{NG}}
\delta \left(\ea-z_2 \left( \frac{\theta_2}{R_0}\right)^\alpha \right),
\end{equation}
where $\Omega_2$ is the (azimuthally averaged) angular function (see for example~\cite{Dokshitzer:1992ip})
\begin{equation} \label{omega2}
\Omega_2 = \frac{2}{\left(1-\cos \theta_1\right)\left(1+\cos \theta_2\right) |\cos \theta_1-\cos \theta_2|}\simeq
 \frac{4}{\theta_1^2  (\theta_1^2- \theta_2^2)}.
\end{equation}
It is now relatively easy to evaluate \Eq{ng-start}. For definiteness, we consider $\beta\ge0$ and obtain
\begin{multline} \label{ng-result}
\frac{\ea}{\sigma}\frac{d \sigma^\text{NG}}{d \ea} = 4 C_F C_A \left ( \frac{\alpha_s}{2 \pi} \right)^2 
\int_{R_0^2}^1 d \theta_1^2 \int_0^{R_0^2} d \theta_2^2 \, \Theta \left(\theta_2^\alpha - R_0^\alpha \ea   \right)
\Theta \left( R_0 \left(\frac{\ea}{\zcut}\right)^{\frac{1}{\alpha+\beta}}- \theta_2  \right)\\ \times
\frac{1}{\theta_1^2  (\theta_1^2- \theta_2^2)} 
\log \frac{\theta_2^\alpha}{R_0^\alpha \ea}\\
=  4 C_F C_A \left ( \frac{\alpha_s}{2 \pi} \right)^2 \Bigg [  {\rm Li}_2\left(\left(\frac{\ea}{\zcut}\right)^\frac{2}{\alpha+\beta}\right) \frac{\alpha \log \frac{1}{\zcut}+\beta \log\frac{1}{ \ea}}{\alpha+\beta} \\+\frac{\alpha}{2}{\rm Li}_3\left( \ea^\frac{2}{\alpha} \right)- \frac{\alpha}{2} {\rm Li}_3\left( \left(\frac{\ea}{\zcut}\right)^\frac{2}{\alpha+\beta} \right) \Bigg]+\order{R_0^2}.
\end{multline}

By itself, \Eq{ng-result} is not particularly enlightening, so it is instructive to take the no grooming limit ($\beta \to \infty$) and the mMDT limit ($\beta=0$). To get a sensible result, we first take the limit of \Eq{ng-result} with respect to $\beta$ and then consider the behavior of the resulting expression at small $\ea$.  For $\beta\to\infty$,
\begin{equation}\label{ng-antikt}
\lim_{\beta \to \infty}\frac{\ea}{\sigma}\frac{d \sigma^\text{NG}}{d \ea} =  C_F C_A \left ( \frac{\alpha_s}{2 \pi} \right)^2 \left( \frac{2}{3}\pi^2 \log \frac{1}{\ea} + \cdots \right) +\order{\beta^{-1}},
\end{equation}
where the dots indicate terms that are not logarithmically enhanced at small $\ea$.  \Eq{ng-antikt} is precisely the result for anti-$k_t$ jets in the small jet radius limit~\cite{BDKM}, and extends to all $\alpha > 0$ since the non-global logarithms arise from soft wide-angle emissions for which the specific angular exponent is a power correction.

For $\beta=0$, there are no non-global logarithms.  In particular, the $\log \ea$ term in \Eq{ng-result} has null coefficient and, after taking the small $\ea$ limit, we obtain
\begin{equation}\label{ng-mMDT}
\lim_{\beta \to 0}\frac{\ea}{\sigma}\frac{d \sigma^\text{NG}}{d \ea}=  C_F C_A \left ( \frac{\alpha_s}{2 \pi} \right)^2  \left(\frac{\ea}{\zcut} \right)^\frac{2}{\alpha}\left( 4 \log \frac{1}{\zcut}- 2\alpha \left(1-\zcut^\frac{2}{\alpha} \right) + \cdots \right) + \,\order{\beta}.
\end{equation}
This expression is consistent with the small-$\zcut$ and small-$R_0$ limit of result for the mMDT mass distribution ($\alpha=2)$~\cite{taggersNLO}.

In general, for finite values of $\beta > 0$, the non-global logarithms are suppressed by powers of $\ea$ with respect to the anti-$k_t$ ($\beta\to \infty$) case.  Taking the small $\ea$ limit of \Eq{ng-result}, we find\footnote{Note that the limits $\beta\to \infty$ and $\ea \to 0$ do not commute with one another as \Eq{ng-antikt} does not follow from the $\beta\to \infty$ limit of \Eq{ng-mMDT-finb}.}
\begin{equation}\label{ng-mMDT-finb}
\lim_{\ea \to 0}\frac{\ea}{\sigma}\frac{d \sigma^\text{NG}}{d \ea}=  4C_F C_A \left ( \frac{\alpha_s}{2 \pi} \right)^2 \frac{\beta}{\alpha+\beta}\left(  \frac{\ea}{\zcut} \right)^{\frac{2}{\alpha+\beta}} \log\frac{1}{\ea}+ \,{\cal O}\left(\ea^{\frac{2}{\alpha+\beta}}\right) \ .
\end{equation}
Because the non-global logarithms are formally power suppressed, we can consistently neglect their resummation to NLL accuracy.  As expected, soft drop declustering removes soft divergences, and hence removes non-global logarithms.

\subsection{Comparison to Monte Carlo}
\label{sec:ang-MC}

We conclude our discussion of $\ea$ by comparing our analytic MLL calculation in \Sec{sec:MLL} (plus the multiple-emission corrections from \Sec{sec:pme}) to a standard Monte Carlo parton shower.  For these simulations, we use \pythia{8.175}~\cite{pythia8} ($p_t$-ordered shower) with the default 4C tune \cite{Corke:2010yf}. We consider proton-proton collisions at $14$~TeV at parton level, including initial- and final-state showering but without multiple parton interactions (i.e.\ UE).  We discuss UE and hadronization corrections in \Sec{sec:NP}.

Jets clustering is performed with the anti-$k_t$ algorithm \cite{anti-kt} with radius $R_0=1.0$~\footnote{We choose $R_0=1.0$ primarily to ease the comparison with previous studies of mMDT in~\Ref{taggersRES}. While we take the small jet radius approximation in this paper, it is known to be reasonable even up to $R_0\sim1$~\cite{Dasgupta:2007wa,DKMS}.} using a development version of \fastjet~3.1 (which for the features used here behaves identically to the 3.0.x series \cite{fastjet}). 
A transverse momentum selection cut $p_T>3$~TeV is applied on the jets before grooming.  To implement the soft drop procedure described in \Sec{sec:softdrop}, jets are reclustered using exclusive C/A \cite{Wobisch:1998wt,Dokshitzer:1997in} to return the same jet.  The soft drop code will be made available as part of the \fastjet\ contrib project (\url{http://fastjet.hepforge.org/contrib/}).

\begin{figure}[]
\begin{center}
\subfloat[]{
\includegraphics[width=7cm]{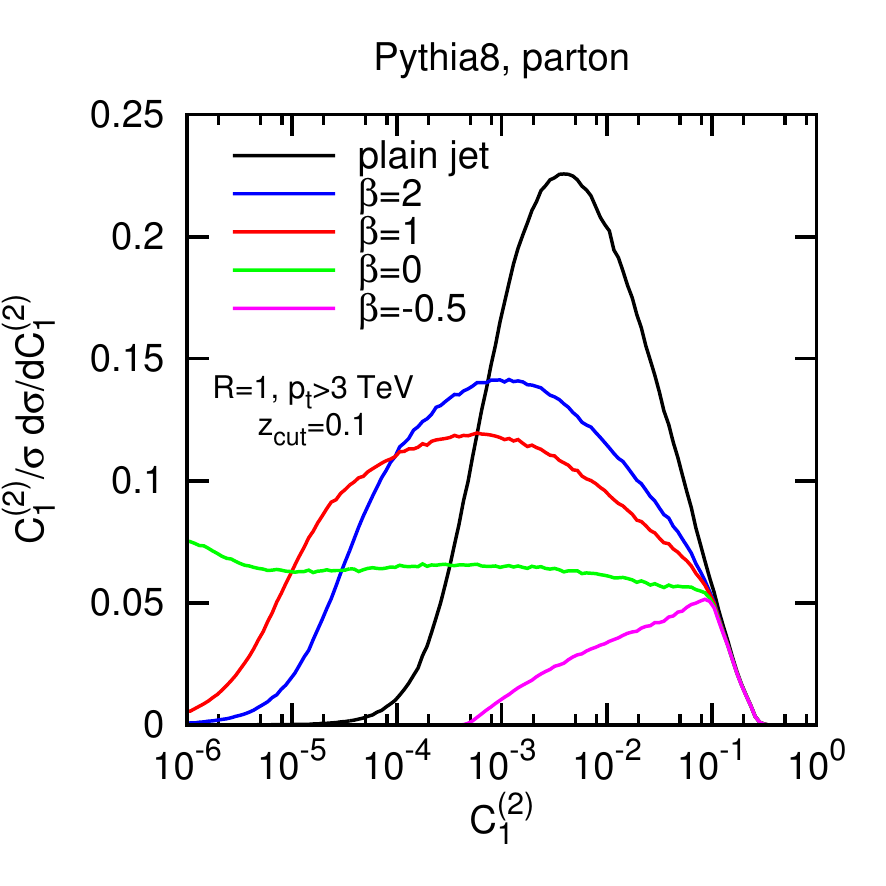}
}
$\quad$
\subfloat[]{\label{fig:sd_mass_MLL}
\includegraphics[width=7cm]{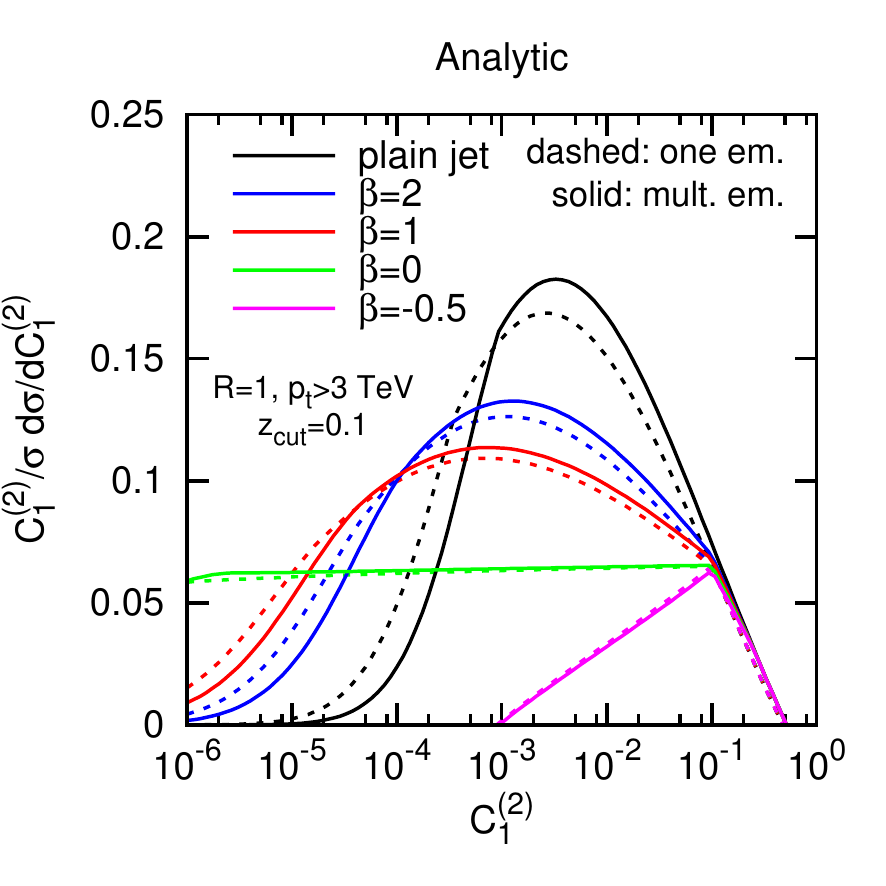}
}
\end{center}
\caption{
The energy correlation functions $C_1^{(\alpha=2)}$ for quark-initiated jets.  Here we compare \pythia{8} \cite{pythia8} (left), our MLL formula in \Eq{all-order-ang-end} (right, dashed curves), and our MLL plus multiple-emissions formula in \Eq{eq:multemis_c1} (right, solid curves).  These $\alpha = 2$ curves correspond to the case of jet mass-squared (normalized to jet energy squared).  We show both the ungroomed (plain jet) distribution, as well as groomed distributions from soft drop declustering with $\zcut = 0.1$ and various values of $\beta$.  For $\beta = 2,1$, we see the expected Sudakov double logarithmic peaks, while $\beta = 0$ (mMDT) has only single logarithms and $\beta = -1$ cuts off at small values.  The \pythia{8} distributions do not have hadronization effects, and the MLL distributions are evaluated by freezing $\alpha_s$ in the infrared.
}
\label{fig:sd_mass}
\end{figure}

We start by considering the case $\alpha=2$, which corresponds to the familiar case of the jet mass distribution. In \Fig{fig:sd_mass} we show results for $qq \to qq$ scattering for different values of angular power $\beta$ in the soft-drop declustering procedure ($\beta=0$ is the mMDT  already studied in \Ref{taggersRES}). The plot on the left has been obtained from \pythia{8}, while the one on the right has been obtained with the analytic resummation, evaluated numerically by freezing the strong coupling in the infrared (see \App{app:angdist}). Dashed curves correspond to MLL accuracy \Eq{all-order-ang-end}, while solid ones include the multiple-emission effect from \Eq{eq:multemis_c1}.

\begin{figure}[p]
\begin{center}
\subfloat[]{
\includegraphics[width=5.7cm]{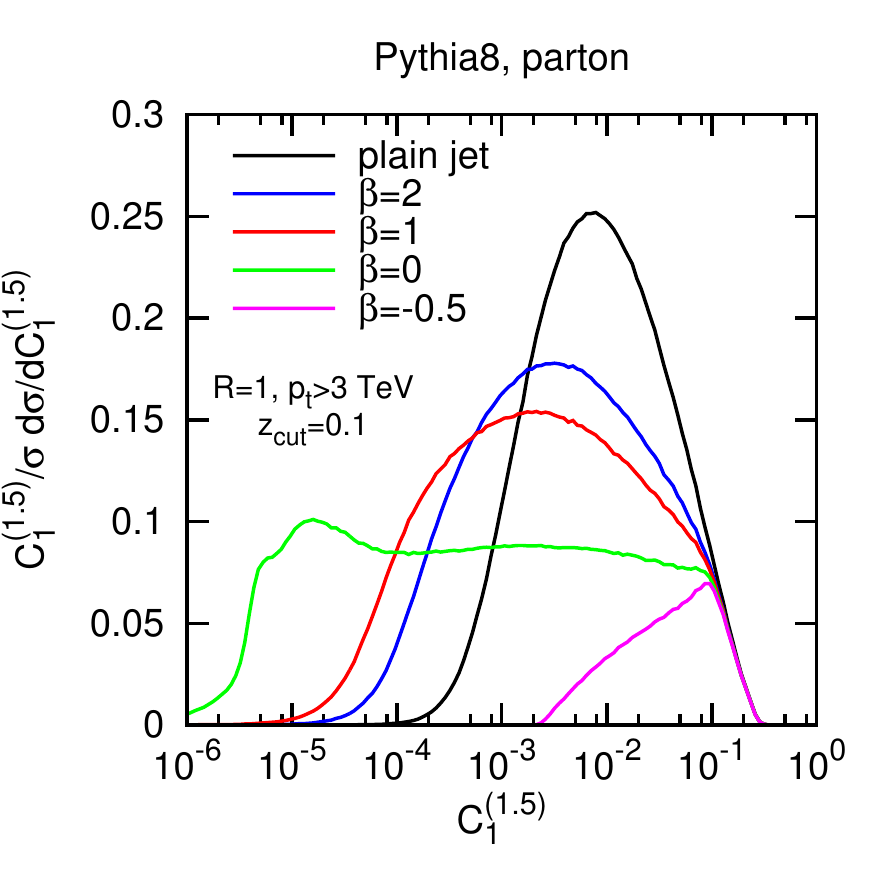}
}
$\quad$
\subfloat[]{
\includegraphics[width=5.7cm]{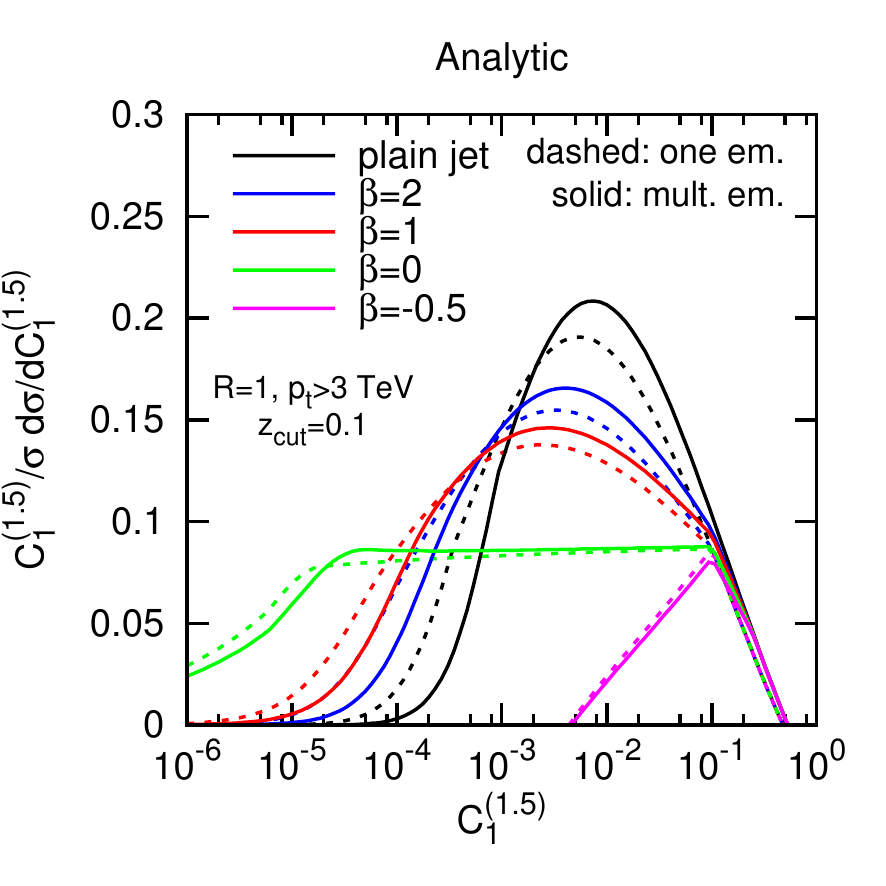}
}
~\\
\subfloat[]{
\includegraphics[width=5.7cm]{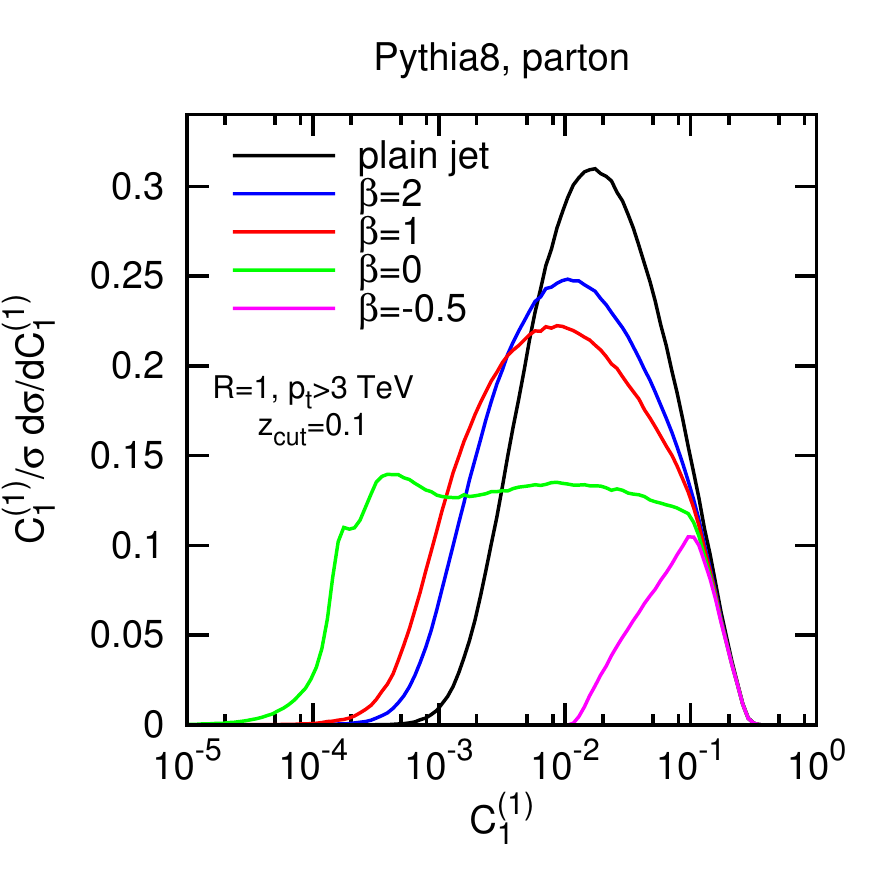}
}
$\quad$
\subfloat[]{
\includegraphics[width=5.7cm]{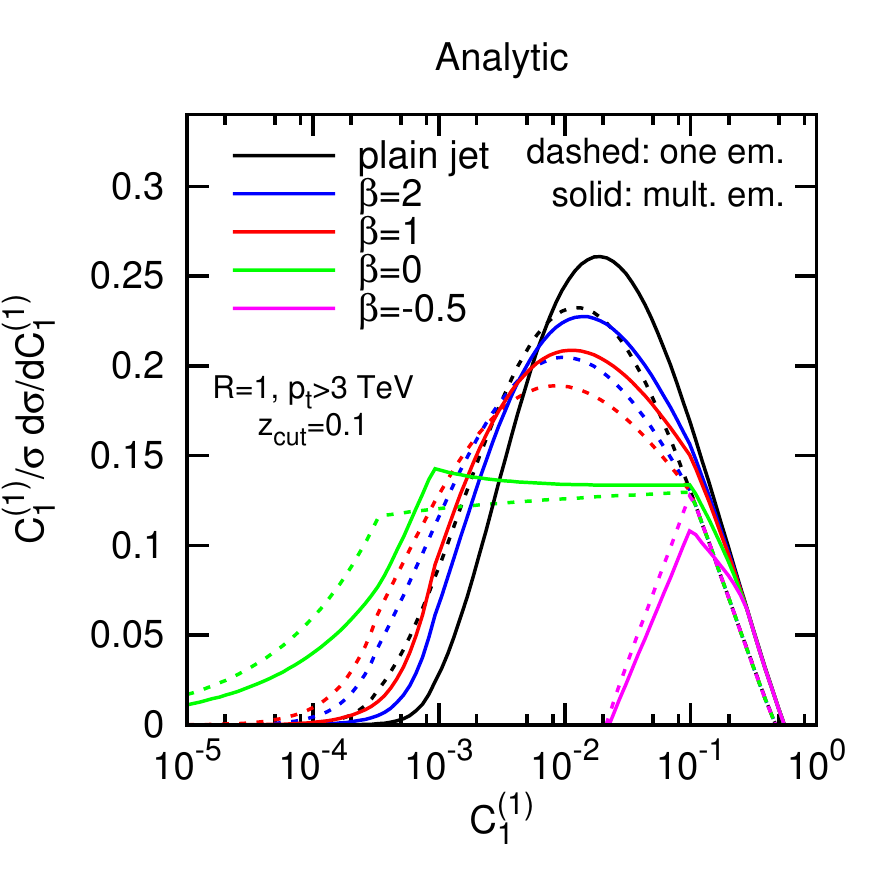}
}
~\\
\subfloat[]{\label{fig:C1alpha05pythia}
\includegraphics[width=5.7cm]{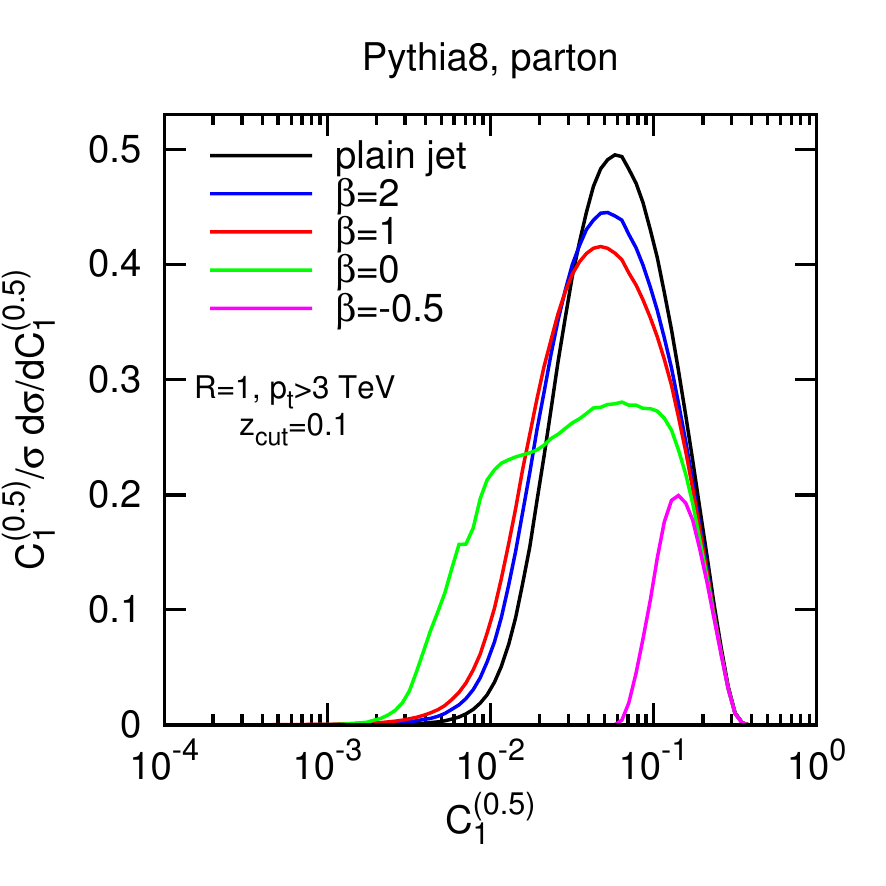}
}
$\quad$
\subfloat[]{\label{fig:C1alpha05analytic}
\includegraphics[width=5.7cm]{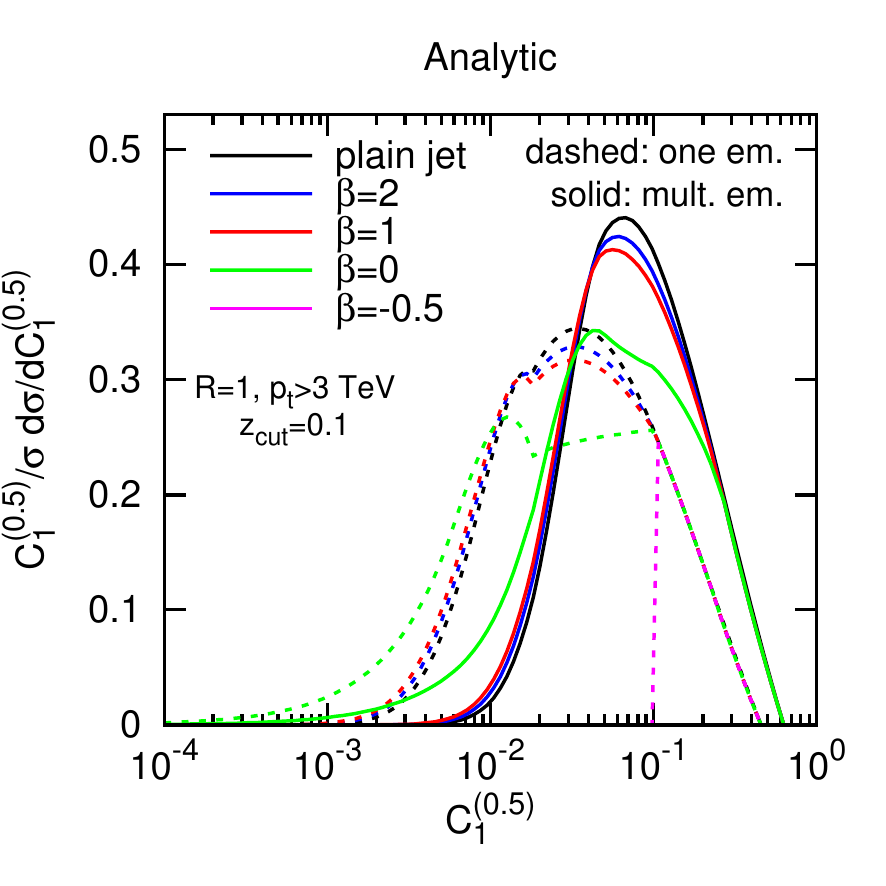}
}
\end{center}
\caption{
The energy correlation functions $\ea$ with $\alpha=1.5,1,0.5$ (top to bottom) for quark-initiated jets.
The plots on the left are obtained with \pythia{8}, while the ones of the right are our MLL predictions (dashed) with multiple emissions included (solid).
}
\label{fig:sd_angsmore}
\end{figure}

The plain jet mass case ($\beta \to \infty$, shown in black) exhibits the characteristic Sudakov peak. All the other curves exhibit a transition point at $\ea=\zcut$ and soft drop is active for $\ea<\zcut$. Soft-dropped distributions with $\beta>0$ (blue and red) are double logarithmic and indeed we can recognize this behavior in the shape of the distribution (i.e.~an upside-down parabola on a log-linear plot).  The case $\beta=0$ (mMDT, green) has no soft logarithms, so the resulting distribution is single logarithmic.  The distribution is nearly flat because the choice $\zcut=0.1$ is close to the value that minimizes higher-order corrections for quark-initiated jets \cite{taggersRES}.  As discussed in \Eq{eq:negbetaminvalue}, the choice of negative $\beta$ (here $\beta=-0.5$ in pink) leads to a distribution with a minimum allowed value, thus regulating both soft and collinear divergencies.

For the groomed distributions, there is good agreement between the parton shower and our analytics.  
Moreover, we also note that the impact of multiple emissions, i.e.\ the difference between solid and dashed curves in \Fig{fig:sd_mass_MLL}, decreases with $\beta$.
It is perhaps surprising that we find worst agreement between analytics and Monte Carlo in the ungroomed (plain jet) case.  However, one should keep in mind that although the two approximations are roughly of the same accuracy (MLL), Monte Carlo parton showers also partially contain many subleading effects. Using the results of \Refs{BDKM,DKMS}, we have checked that subleading effects (like initial-state radiation and non-global logarithms) play a non-negligible role.  Indeed, \pythia{8} is closer to the full NLL result than to the (less accurate) MLL plus multiple emissions one presented here.  Because the action of soft drop is to remove large-angle soft radiation (e.g.\ initial state radiation and non-global logarithms), it is reassuring that our calculations for the finite $\beta$ soft-drop curves are indeed in better agreement with the parton shower.

In \Fig{fig:sd_angsmore}, we compare our analytic resummation to the parton shower for $\ea$ with $\alpha=1.5,1,0.5$.  Again, the plots on the left are obtained with \pythia{8} while the ones on the right are the MLL plus multiple emissions results.   The same gross features seen with $\alpha = 2$ are also present here, including the fact that the agreement between Monte Carlo and analytics is better with grooming than without.   Overall, however, the agreement gets worse as $\alpha$ decreases.  This is likely because, as seen in \Eq{ang-LO-res}, the expansion parameter is really $\alpha_s/\alpha$, so both logarithmically-enhanced and non-singular fixed-order corrections are more important at small $\alpha$.  It is encouraging that the peak locations are still roughly the same in the analytic calculations and \pythia{8} results, even if the overall peak normalizations slightly differ.
We note that the dashed curves in \Fig{fig:C1alpha05analytic} have kinks; indeed all the curves in this section obtained from analytic calculations suffer from the same behavior, although this feature is not visible on the other plots. The position of the kink is 
$\ea =\left(\frac{\mu_\text{NP}}{p_T R_0} \right)^\alpha$
 and it is a consequence of the way we freeze the running coupling at $\kappa=\mu_\text{NP}$. As detailed in \App{app:angdist}, this effect is beyond the accuracy of our calculation.
Finally,  in \Figs{fig:C1alpha05pythia}{fig:C1alpha05analytic}, note the sharp cutoff of the plots when $\alpha + \beta = 0$, which can be understood from \Eq{eq:negbetaminvalue}.
In \Fig{fig:C1alpha05analytic}, we only show the MLL result since fixed-order corrections are expected to be important and our treatment of multiple emissions effects in \Eq{finite-derivative} becomes singular when $\alpha+\beta=0$.

\section{Groomed Jet Radius}
\label{sec:pileup}

Because the soft drop procedure is defined through declustering a C/A branching tree, there is a well-defined and IRC-safe meaning to the groomed jet radius.  Concretely, the groomed jet radius $R_g$ is the angle between the branches that first satisfy \Eq{eq:mainrelation}, which is sensible because for a C/A tree, all subsequent branches are separated by an angle less than $R_g$.  From a practical perspective, $R_g$ is particularly interesting, since the groomed jet area is approximately $\pi R_g^2$.  Thus, $R_g$ serves as a proxy for the sensitivity of the groomed jet to possible contamination from pileup \cite{Cacciari:2008gn,Sapeta:2010uk}.

\subsection{Modified Leading Logarithmic Approximation}
\label{sec:MLLradius}

\begin{figure}
\begin{center}
\includegraphics[scale=0.55]{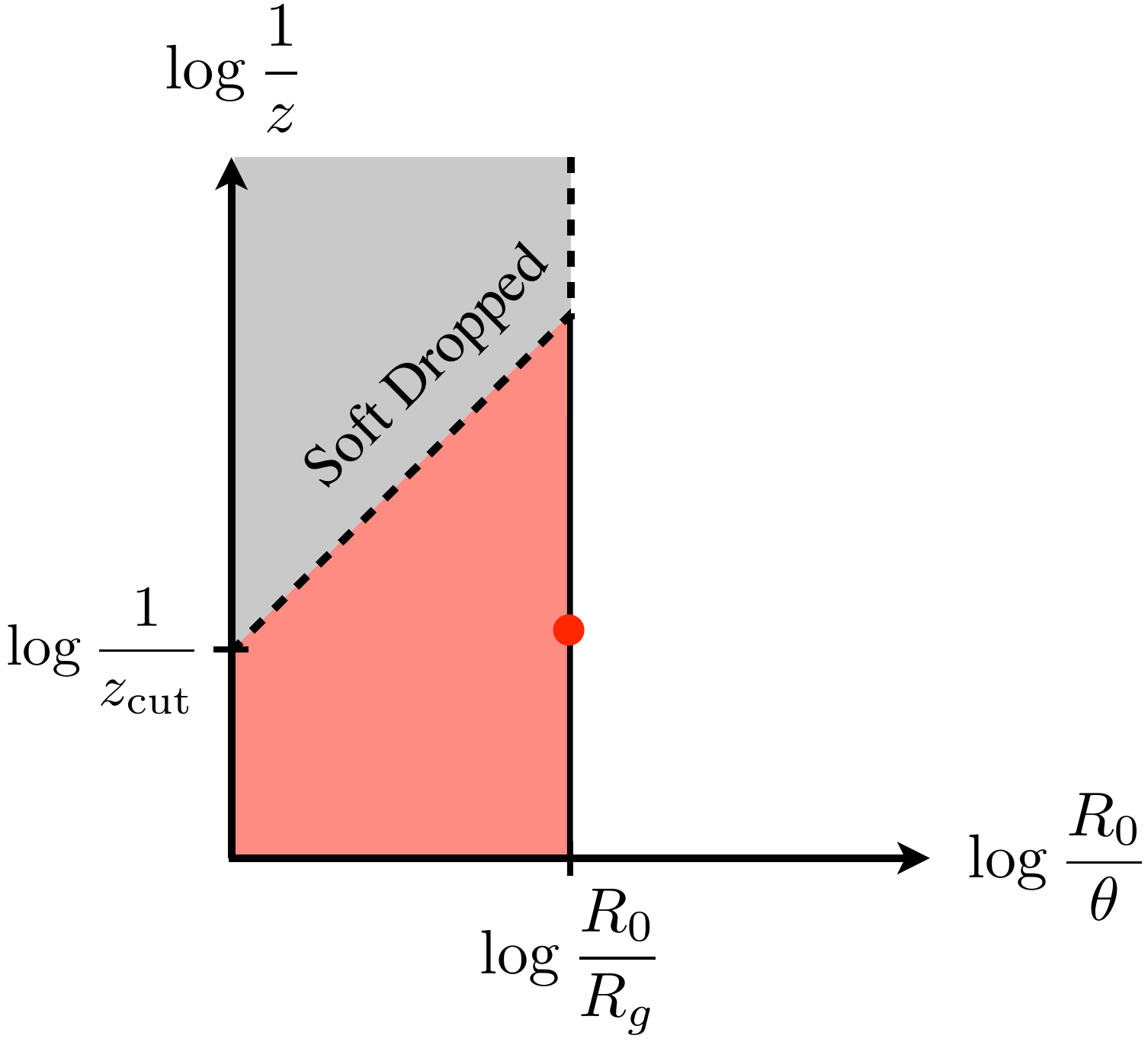}
\end{center}
\caption{Phase space for emissions relevant for groomed jet radius $R_g$ in the $(\log \frac{1}{z},\log \frac{R_0}{\theta})$ plane.  The soft dropped region is gray and the first emission satisfying the soft drop criteria is illustrated by the red dot.  The forbidden emission region (the Sudakov exponent) is shaded in pink.
}
\label{fig:groomedregions-radius}
\end{figure}

The calculation of the groomed jet radius to MLL accuracy follows much of the same logic as the $\ea$ calculation in \Sec{sec:MLL}.  As illustrated in \Fig{fig:groomedregions-radius}, $R_g$ actually corresponds to a simpler phase space than $\ea$.  A given value of $R_g$ simply means that all emissions at angles larger than $R_g$ failed the soft drop criteria.  Therefore, the  $R_g$ distribution can be calculated by demanding that all emissions at angles larger than $R_g$ were groomed away.
As explained in more detail in \App{app:rdist}, this understanding translates into the following cumulative distribution for the groomed jet radius:
\begin{align}\label{eq:grad_exp}
\Sigma^\text{radius}(R_g) &= \exp\left[
-\int_{R_g}^{R_0} \frac{d\theta}{\theta}\int_0^1 dz\, p_i(z) \, \frac{\alpha_s(\kappa)}{\pi}\Theta\left( z- z_\text{cut}\frac{\theta^\beta}{R_0^\beta} \right)
\right] \ ,
\end{align}
where the integral in the exponent again corresponds to vetoed emissions (i.e.\ the pink region in \Fig{fig:groomedregions-radius}).

Besides the simpler phase space for one emission, $R_g$ is also simpler than $\ea$ with respect to multiple emissions.  In the case of $\ea$, multiple emissions could contribute to the value of $\ea$, but the MLL approximation effectively only considers the contribution from a single dominant emission.  For $R_g$, though, once one emission satisfies the soft drop criteria, the jet radius is set, so multiple emissions do not contribute to this observable. We have also verified that non-global contributions are suppressed by $R_g$ for $\beta<\infty$, analogously to the energy correlation case. For these reasons, we believe that the expression in \Eq{eq:grad_exp} is fully accurate to single-logarithmic level,\footnote{Strictly speaking, NLL accuracy requires evaluating the strong coupling at two loops, i.e.\ with $\beta_1$, in the CMW scheme \cite{Catani:1990rr}.} though for consistency with the rest of this paper, we will only evaluate  \Eq{eq:grad_exp}  to MLL accuracy.

\subsection{Comparison to Monte Carlo}
\begin{figure}[]
\begin{center}
\subfloat[]{\label{fig:py_grrad}
\includegraphics[width=7cm]{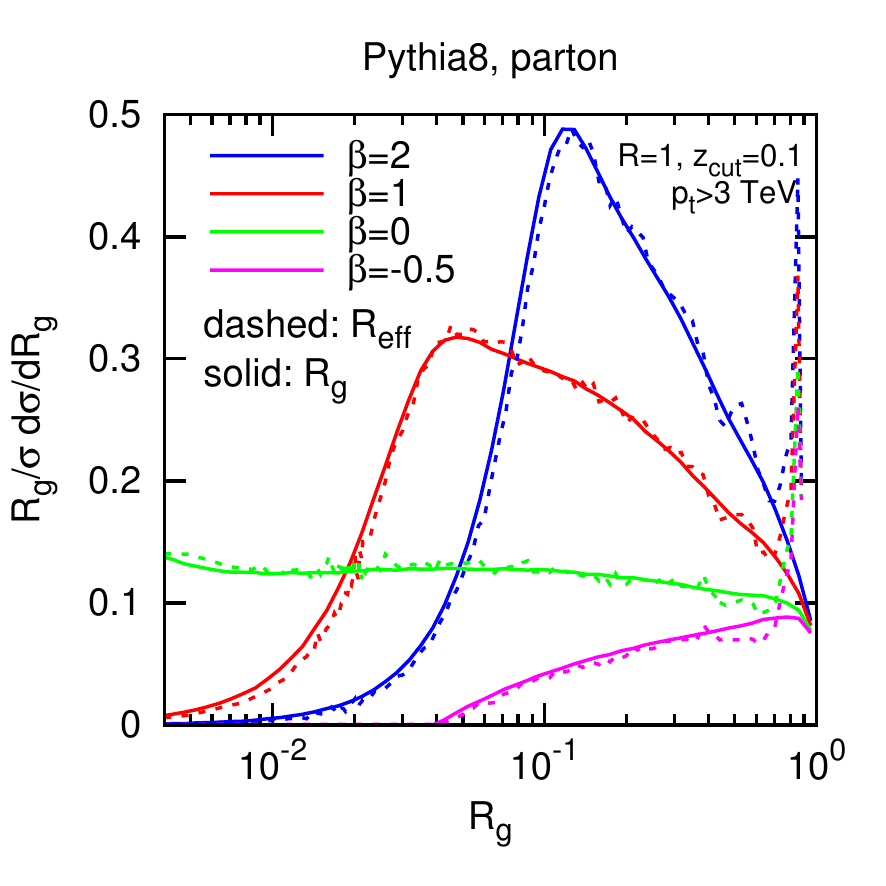}
}
$\quad$
\subfloat[]{\label{fig:anal_grrad}
\includegraphics[width=7cm]{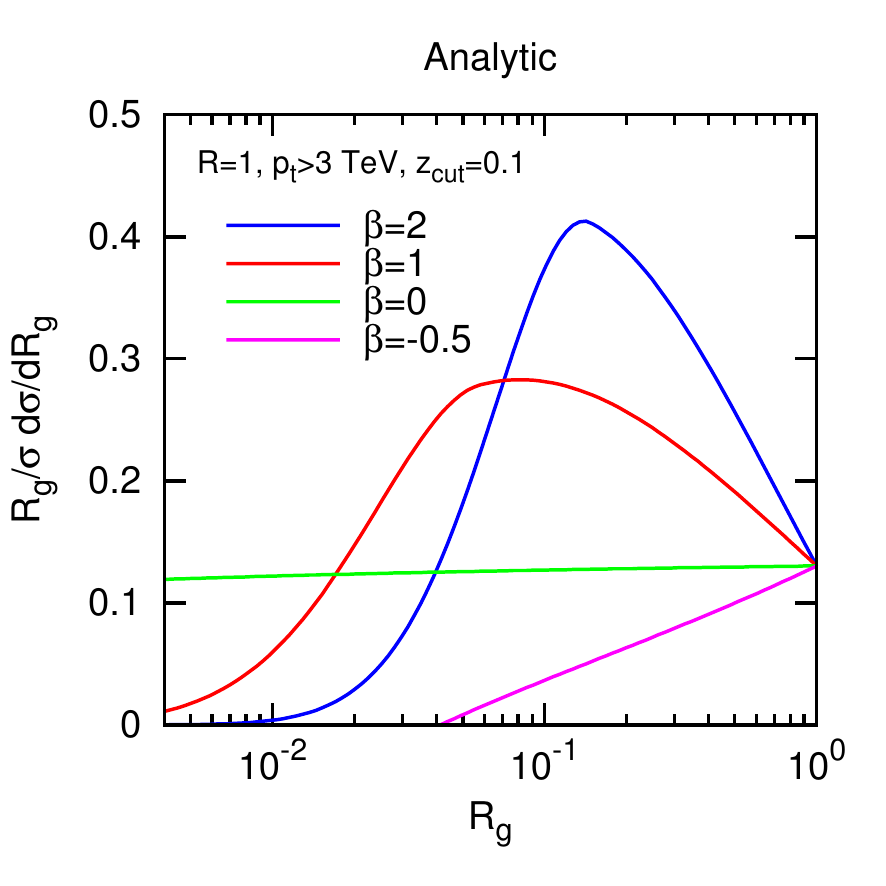}
}

\end{center}
\caption{ Comparison of the jet radius $R_g$ distribution extracted
  from \pythia{8} (left, solid) and inferred from the active area in
  \pythia{8} using $R_\text{eff} =\sqrt{A_\text{active}/\pi \xi}$
  (left, dashed), and computed to MLL accuracy (right).  The original
  jet radius is set to be $R_0 = 1$ and the jets have an ungroomed
  energy of $3$ TeV.  The soft drop parameter is $z_\text{cut}=0.1$,
  while $\beta$ is varied.  }
\label{fig:rad_dist}
\end{figure}

There are two different ways one can define the groomed jet radius in
Monte Carlo. The first method is to simply measure the $R_g$ value of
the C/A branching that satisfies the soft drop condition.  A second
approach, more directly relevant for pileup mitigation, is to
determine the effective radius of the groomed jet from its active area
\cite{jet-area}.  The active area of a jet is defined as the area over
which infinitesimally soft particles are clustered into the jet. An
effective jet radius $R_\text{eff}$ can then be defined from the
groomed jet active area using:
\begin{equation}
\label{eq:Reffdef}
R_\text{eff} \equiv \left(\frac{A_\text{active}}{\pi \xi }\right)^{1/2} \ ,
\end{equation}
where $A_\text{active}$ is the active jet area, and $\xi \simeq
(1.16)^2$ accounts for the fact that a typical C/A jet of radius
$R_0$ has an average active area $\xi \pi R_0^2$.\footnote{The
  numerical value for $\xi$ can be read from Fig. 8 in
  \Ref{jet-area}. Strictly speaking, this result is only valid for a jet made of two
  particles separated by $R_0$, with one of them much softer than the other. However, for
  C/A jets, one expects that this would not vary much for more
  symmetric two-particle configurations (see e.g.~\Ref{Sapeta:2010uk}).} 

In \Fig{fig:py_grrad} we show the $R_g$ and $R_\text{eff}$ distributions as measured on the same \pythia{} jet samples introduced in \Sec{sec:ang-MC}. To obtain $R_\text{eff}$ in practice, we have computed the groomed jet area using active areas as implemented in \fastjet~(v3), and we used a ghost area of 0.0005 and 10 repetitions in order to reach sufficiently small values of $R_{\rm eff}$.  With the $\xi$ offset factor, the two techniques give remarkably similar results, giving strong evidence that the groomed jet radius $R_g$ is an effective measure of pileup sensitivity.  The main difference is the spike at $R_{\rm eff} = 1/\sqrt{\xi}$, corresponding to cases where the first C/A branching already satisfies the soft drop condition, yet typically with $R_g < 1$.  The nice reduction of the jet area even with mild grooming (e.g.~$\beta = 2$) suggests that soft drop should work well for pileup mitigation, but we leave a detailed study to future work.

In \Fig{fig:anal_grrad}, we show the MLL distribution from \Eq{eq:grad_exp}.  There is good qualitative agreement with \pythia{} for a range of angular exponents $\beta$, suggesting that our MLL calculation for $R_g$ captures the relevant physics effects present in the Monte Carlo simulation.

\section{Jet Energy Drop}
\label{sec:energy}

\begin{figure}
\begin{center}
\subfloat[]{
\includegraphics[width=7cm]{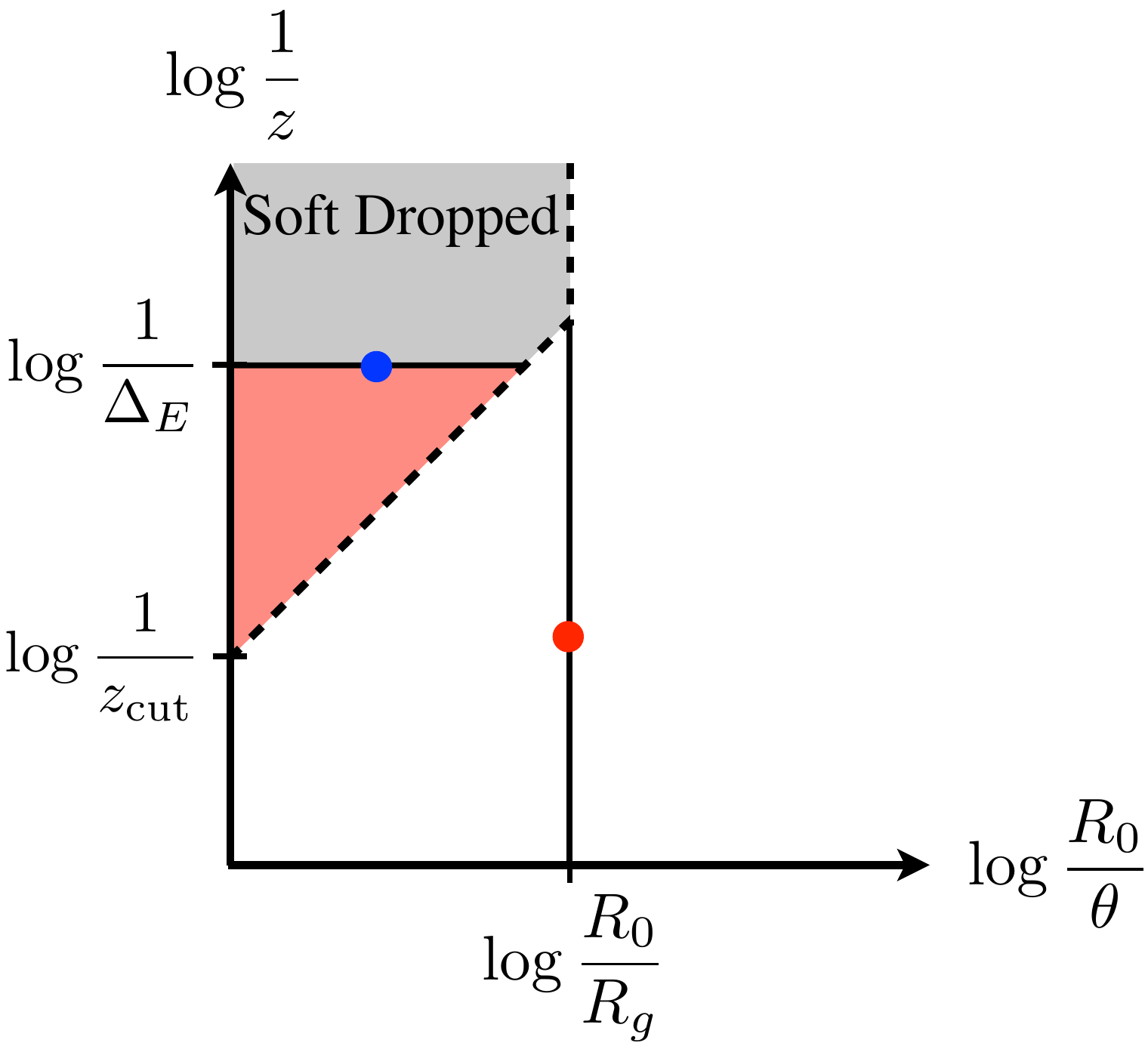}
}
$\quad$
\subfloat[]{
\includegraphics[width=7cm]{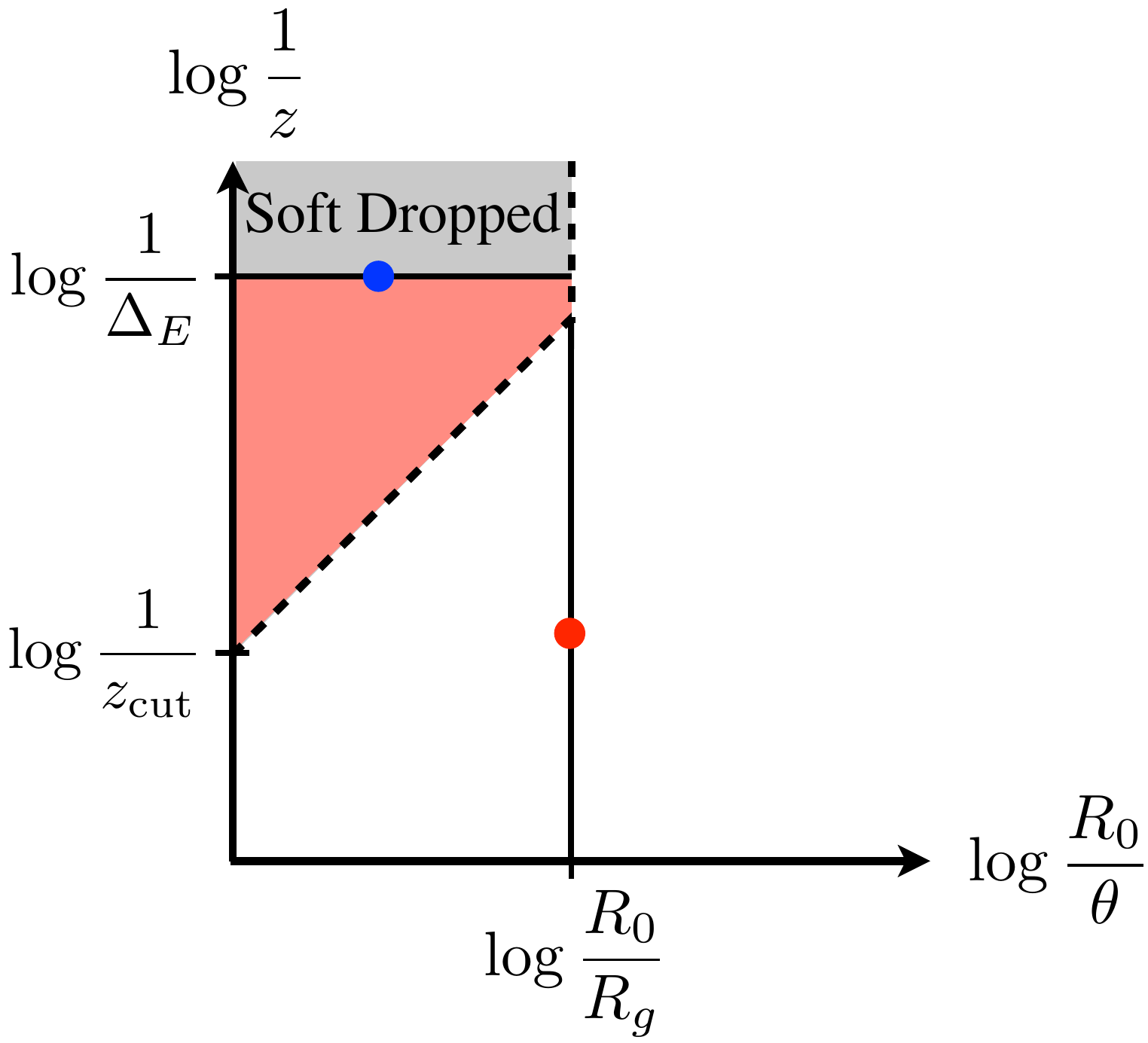}
}
\end{center}
\caption{Phase space for emissions relevant for groomed jet energy loss $\Delta_E$ in the $(\log \frac{1}{z},\log \frac{R_0}{\theta})$ plane.  The soft dropped region is gray/pink and the first emission satisfying the soft drop criteria is illustrated by the red dot, located at the groomed jet radius, $R_g$.  The blue dot represents the leading contribution to $\Delta_E$, with subleading contributions above it.  The location in angle of all soft dropped emissions is larger than $R_g$.  The forbidden emission region for a given value of $R_g$ (the Sudakov exponent) is shaded in pink.  The left (right) plot shows $\Delta_E$ larger (smaller) than $\zcut (R_g/R_0)^\beta$.
}
\label{fig:groomedregions-energy-loss}
\end{figure}

Our final analytic calculation is for the groomed jet energy.  Unlike for many other grooming procedures, the energy of a soft-drop jet ($\beta>0$) is an IRC safe observable and so can be computed in pQCD.  In particular, we will study the fractional energy drop due to grooming $\Delta_E$ defined as
\begin{equation}
\Delta_E \equiv \frac{E_0 - E_g}{E_0} \ ,
\end{equation}
where $E_0$ is the energy of the jet before grooming and $E_g$ is the energy of the groomed jet.  $\Delta_E$ can be interpreted as a measure of the fraction of the original jet's energy contained in soft wide-angle emissions.  In the small $R_0$ limit, $\Delta_E$ is the same as the fractional $p_T$ loss, which is the more relevant quantity for non-central ($y \not= 0$) jets in hadronic collisions.

\subsection{Modified Leading Logarithmic Approximation}
\label{sec:MLLenergy}

At MLL order, the calculation of the $\Delta_E$ distribution is more subtle than for $\ea$ or $R_g$.  In the case of $\ea$ and $R_g$, the Sudakov veto region was effectively determined by a single emission, and the multiple emissions effect for $\ea$ could be included as a higher-order correction (see \Sec{sec:pme}).  

For $\Delta_E$, the veto region depends crucially on two emissions, as illustrated in \Fig{fig:groomedregions-energy-loss}.  The energy drop due to grooming comes from large angle emissions that fail the soft drop condition.  But to figure out which emissions are dropped, we first have to know which emission satisfied the soft drop condition, since that sets the groomed jet radius $R_g$.  All emissions lying at angles greater than $R_g$ are removed by soft drop, but all emissions at angles less than $R_g$ are maintained.  Thus, the energy drop depends both on the emission that sets $R_g$ and on the emissions that contributes to $\Delta_E$.

In practice, the easiest way to determine the $\Delta_E$ distribution is by computing the energy drop for a given value of $R_g$ and then integrating over the $R_g$ distribution.  In equations, the cumulative distribution of $\Delta_E$ is given by
\begin{equation}\label{eq:elossdef}
\Sigma^\text{energy-drop}(\Delta_E) = \int_0^{R_0} dR_g \, \frac{d \Sigma^\text{radius}(R_g)}{d R_g}\, \widetilde{\Sigma}(R_g,\Delta_E) \ ,
\end{equation}
where $\widetilde{\Sigma}(R_g,\Delta_E)$ is the cumulative distribution of $\Delta_E$ for a given groomed jet radius $R_g$.  The cumulative distribution $\Sigma^\text{radius}(R_g)$ was defined in \Eq{eq:grad_exp}, and the derivative factor is needed to extract the differential cross section (i.e.\ the probability distribution) for $R_g$.

The details of the $\widetilde{\Sigma}(R_g,\Delta_E)$ calculation are presented in \App{app:edist}.  The key is that this double cumulative distribution can be computed at logarithmic accuracy by summing over independent contributions at all orders:
\begin{align}
\widetilde{\Sigma}(R_g,\Delta_E)&=\sum_{n=1}^\infty \left[  \prod_{m=1}^n \int_{R_g}^{R_0} \frac{d\theta_m}{\theta_m} \int_0^1 dz_m \, p_i(z_m) \frac{\alpha_s(\kappa_{m})}{\pi} \Theta(\theta_m-\theta_{m+1})  \Theta\left( z_\text{cut}\frac{\theta_m^\beta}{R_0^\beta}-z_m  \right)\right]
\nonumber \\
&\qquad\quad
\times\Theta\left( \Delta_E - \sum_{m=1}^n z_m \right)\exp\left[-\int_{R_g}^{R_0}\frac{d\theta}{\theta} \int_0^1 dz\, p_i(z) \frac{\alpha_s(\kappa)}{\pi} \Theta\left( z_\text{cut}\frac{\theta^\beta}{R_0^\beta}-z  \right) \right] \nonumber \\
&=\int \frac{d\nu}{2\pi i \nu}e^{\nu \Delta_E}  e^{-R_2\left(R_g,\nu^{-1}\right)} \ .
\end{align}
Here, we are accounting for the effect of multiple emissions (i.e.\ the sum over $m$ in the observable) by performing a Laplace transform, and the explicit integral over $\nu$ represents the inverse Laplace transform.  
The radiator function appearing in the exponent is 
\begin{equation}\label{eq:elossrad}
R_2\left(R_g,\nu^{-1}\right)=\int_{R_g}^{R_0}\frac{d\theta}{\theta} \int_0^1 dz\, p_i(z) \frac{\alpha_s(\kappa)}{\pi}\Theta\left( z_\text{cut}\frac{\theta^\beta}{R_0^\beta}-z  \right)\left( 1- e^{-\nu z}  \right) \ ,
\end{equation}
which is a function of both the Laplace transform parameter $\nu$ and the groomed jet radius $R_g$.

\subsection{Sudakov Safety for $\beta=0$}
\label{sec:sdkv}

As mentioned in \Sec{sec:betachange}, the groomed jet energy drop $\Delta_E$ is IRC safe only if $\beta>0$.  In particular, the energy of a $\beta = 0$ (mMDT) groomed jet is not an IRC safe quantity, since a measured value of energy does not require two well-separated hard prongs in the jet.

On the other hand, \Eq{eq:elossdef} has a smooth $\beta \to 0$ limit, and therefore is still calculable (despite being IRC unsafe).  Specifically, we are calculating the $\Delta_E$ distribution at a fixed groomed jet radius $R_g$, which forces a two-prong configuration.  There is still an (IRC unsafe) singularity at $R_g \to 0$, but this is regulated by the Sudakov factor in the $R_g$ distribution.  This property was referred to as ``Sudakov safety'' in \Ref{Larkoski:2013paa}.  As we will now show, the way in which IRC unsafety but Sudakov safety manifests itself for $\Delta_E$ is rather peculiar.

The behavior of $\Delta_E$ for $\beta=0$ is easiest to study by computing the cumulative distribution of the energy drop at fixed coupling.  We will also take the Laplace conjugate parameter $\nu\to \infty$ to suppress multiple emissions effects.  This limit removes the inverse Laplace transform and turns the exponential factor in \Eq{eq:elossrad} into the constraint that $z>\Delta_E$.  We emphasize that the $\nu\to \infty$ limit is only taken to simplify the following discussion; the fixed-coupling energy loss distribution with the full multiple emissions effect exhibits the same properties.

At fixed-coupling, the cumulative distribution of the groomed jet radius is
\begin{align}
\Sigma^\text{radius}(R_g)& \overset{\rm f.c.}{\wideeq{0.6cm}}
 \exp\left[ -\frac{\alpha_s}{\pi} \int_{R_g}^{R_0} \frac{d\theta}{\theta} \int_{\zcut}^1 dz \, p_i(z)\,\Theta\left( z- z_\text{cut}\frac{\theta^\beta}{R_0^\beta} \right) \right] \nonumber \\
&\simeq \exp\left[ -\frac{\alpha_s}{\pi} C_i\left( \beta \log^2 \frac{R_0}{R_g} - 2 \log \zcut \log \frac{R_0}{R_g}+2 B_i \log \frac{R_0}{R_g} \right) \right]  \ ,
\end{align}
where we have ignored terms suppressed by positive powers of $\zcut$ and $\Delta_E$.
The cumulative distribution of the energy drop at fixed groomed jet radius is
\begin{align}
\widetilde{\Sigma}(R_g,\Delta_E) &
 \overset{\rm f.c.}{\wideeq{0.6cm}}\exp\left[ -\frac{\alpha_s}{\pi} \int_{R_g}^{R_0} \frac{d\theta}{\theta} \int^{\zcut}_{\Delta_E} dz \, p_i(z) \, \Theta\left( z_\text{cut}\frac{\theta^\beta}{R_0^\beta}-z  \right) \right] \nonumber \\
&\simeq 
\Theta\left( \zcut \frac{R_g^\beta}{R_0^\beta} - \Delta_E  \right) \, \exp\left[   
-\frac{\alpha_s}{\pi}C_i \left( 
2\log \frac{\zcut}{\Delta_E} \log \frac{R_0}{R_g} - \beta \log^2\frac{R_0}{R_g}
\right)
\right] \nonumber \\
&\qquad+
\Theta\left( \Delta_E-\zcut \frac{R_g^\beta}{R_0^\beta} \right)\Theta(\zcut - \Delta_E)\, \exp\left[
-\frac{\alpha_s}{\pi}\frac{C_i}{\beta} \log^2\frac{\zcut}{\Delta_E}
\right]
  \ .
\end{align}
Plugging these expressions into \Eq{eq:elossdef} in the $\nu \to \infty$ limit, we find the cumulative distribution of the groomed energy drop to be
\begin{align}\label{eq:fixacumeloss}
\Sigma^\text{energy-drop}(\Delta_E) = 
\frac{\log \zcut - B_i}{\log \Delta_E - B_i}
+\frac{\pi \beta}{2 C_i \alpha_s (\log\Delta_E - B_i)^2} 
\left(1-e^{-2\frac{\alpha_s}{\pi}\frac{C_i}{\beta}\log\frac{\zcut}{\Delta_E}\left(\log \frac{1}{\Delta_E} + B_i\right)}
\right)
\ ,
\end{align}
for $\Delta_E<\zcut$.  At this order, the cumulative distribution is constant for $\Delta_E > \zcut$.

The expression in \Eq{eq:fixacumeloss} has some fascinating properties.  First, by expanding order-by-order in $\alpha_s$, we find 
\begin{equation}
\Sigma^\text{energy-drop}(\Delta_E) = 1-\frac{\alpha_s}{\pi}\frac{C_i}{\beta}\log^2\frac{\zcut}{\Delta_E} + {\cal O}\left(\left(\frac{\alpha_s}{\beta}\right)^2\right) \ .
\end{equation}
Thus, the expansion in powers of the strong coupling is actually an expansion in $\as/\beta$, which diverges order-by-order in perturbation theory for $\beta \to 0$.  Thus, as advertised, the energy drop distribution is not IRC safe for $\beta=0$.
However, the $\beta \to 0$ limit of \Eq{eq:fixacumeloss} can be taken before expanding in $\alpha_s$.  The $\beta \to 0$ limit yields the simple and surprising result
\begin{equation}
\label{eq:energydropnoalpha}
\Sigma^\text{energy-drop}(\Delta_E)_{\beta = 0} = \frac{\log \zcut - B_i}{\log \Delta_E - B_i} \ ,
\end{equation}
which is completely independent of $\alpha_s$!  So while the strong coupling constant $\alpha_s$ was necessary to calculate $\Delta_E$, the leading behavior is independent of the value of $\alpha_s$.

We can attribute this behavior to the fact that $\Delta_E$ is a Sudakov safe observable for $\beta=0$.  The singular region of phase space at $R_g \to 0$ is exponentially suppressed by the Sudakov factor in $\Sigma^\text{radius}(R_g)$.  This exponential suppression balances the exponential increase in the number of groomed emissions in such a way that $\Delta_E$ is independent of $\alpha_s$.  In fact, $\Delta_E$ is independent of the total color of the jet at fixed coupling, and only depends on the flavor of the jet through the subleading terms in the splitting functions $B_i$.   When the running coupling is included, we will see that the dominant contribution to the $\Delta_E$ distribution is still independent of $\alpha_s$, with only weak dependence controlled by the QCD $\beta$-function.

\subsection{Non-Global Logarithms}
\label{sec:NGforDeltaE}

The ungroomed jet energy $E_0$ is clearly affected by non-global contributions, since emissions outside of the jet can radiate energy into the jet.  Because the soft drop procedure removes soft wide-angle radiation, we expect that the groomed jet energy $E_g$ should have no non-global contributions.  In principle, we could calculate the $E_g$ distribution directly to show the absence of non-global logarithms.  In practice, though, it is hard to interpret the meaning of $E_g$ without invoking some reference energy scale.  Here, we are using $E_0$ as a reference, which is not ideal since $E_0$ has non-global contributions.  That said, we will find that the $E_0$ and $\Delta_E$ distributions have exactly the same non-global logarithms, implying that the $E_g$ distribution is wholly absent of them.

Analogous to \Sec{sec:ang-NGLs}, we can do a simple calculation of the non-global contribution to $\Delta_E$.
At lowest order for a narrow jet of radius $R_0$, the non-global logarithms can be computed from
\begin{align}
\frac{1}{\sigma_0}\frac{d\sigma^\text{NG}}{d\Delta_E} &= 4 C_F C_A \left( \frac{\alpha_s}{2\pi} \right)^2 \int \frac{dz_1}{z_1}\frac{dz_2}{z_2} \int d\theta_1 \, d\theta_2 \frac{4 \theta_1\theta_2}{\theta_1^2\left(  \theta_1^2 - \theta_2^2 \right)} \Theta_\text{NG}\Theta\left(\zcut\frac{\theta_2^\beta}{R_0^\beta} - z_2 \right)  \delta(\Delta_E - z_2) \nonumber\\
&=\frac{2}{3}\pi^2 C_F C_A \left( \frac{\alpha_s}{2\pi} \right)^2 \frac{\log\frac{1}{\Delta_E}}{\Delta_E}+{\cal O}\left(R_0^2,\frac{\Delta_E^{2/\beta}}{\zcut^{2/\beta}}\right) \ .
\end{align}
This shows that non-global logarithms are not power-suppressed for the energy loss distribution regardless of $\beta$.  Moreover, the coefficient of the non-global logarithms are the same for the ungroomed distribution ($\beta \to \infty$) as for the groomed distribution (finite $\beta$).  This implies that the groomed jet energy $E_g$ cannot contain any non-global logarithms.  

Of course, to really verify this behavior, one would want to calculate the groomed jet energy distribution in a process with an additional scale.  For example, one could study the associated production of a photon and a jet (i.e.~$p p \to \gamma +j$) and use the photon momentum as a reference scale.  In this example, we would expect $(p_{Tg} - p_{T\gamma}) / p_{T\gamma}$ should be free of non-global logarithms.

\subsection{Comparison to Monte Carlo}

\begin{figure}[]
\begin{center}
\subfloat[]{
\includegraphics[width=7cm]{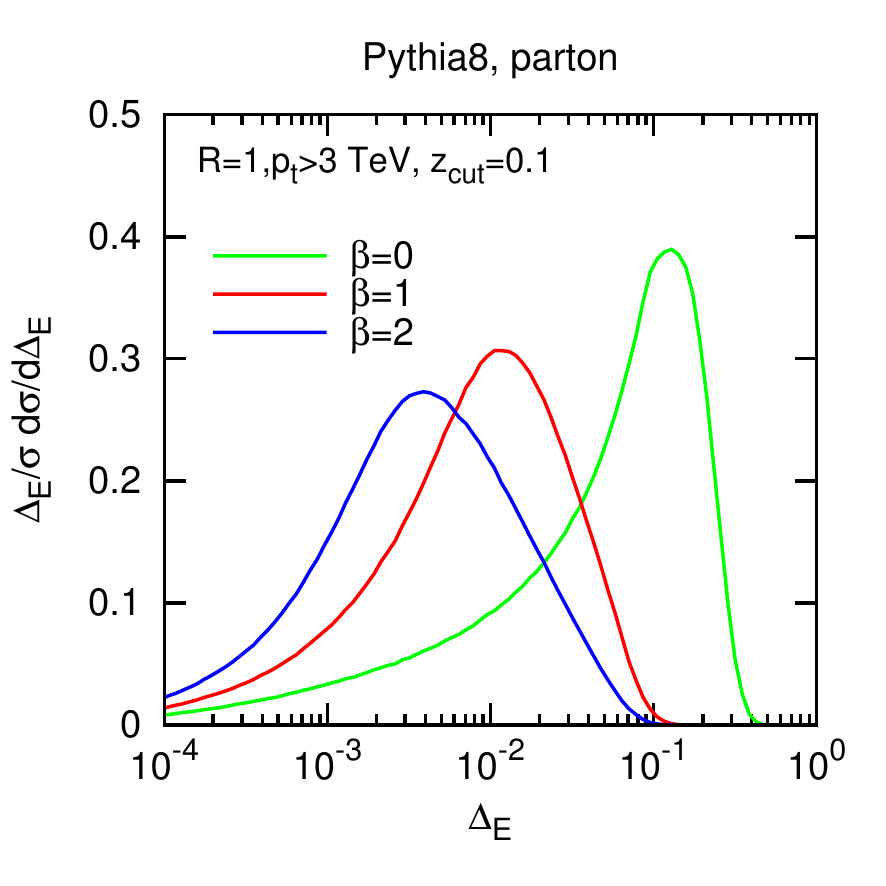}
}
$\quad$
\subfloat[]{
\includegraphics[width=7cm]{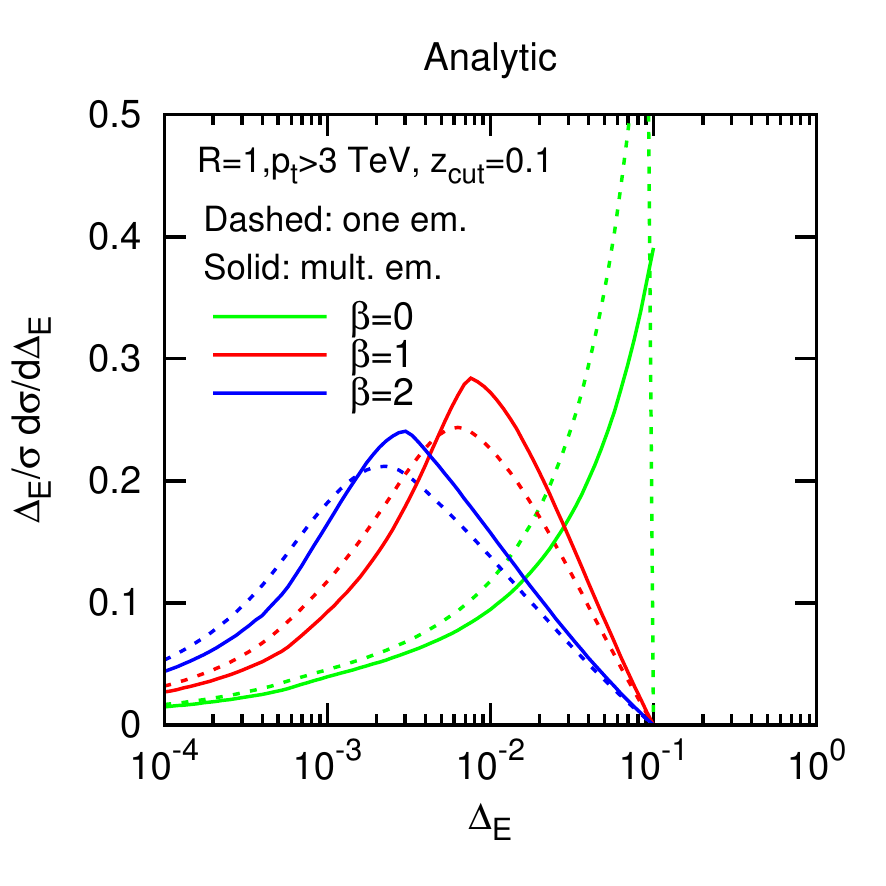}
}
\end{center}
\caption{
The distribution of groomed energy loss $\Delta_E$ in \pythia{8} (left) compared to our MLL calculation (right).  In the MLL result, solid (dashed) corresponds to the distribution with (without) multiple emissions.  The original jet radius is set to  $R_0 = 1.0$ and the jets have an ungroomed energy of $3$ TeV.  The soft drop parameter $z_\text{cut}=0.1$ is fixed while $\beta$ is varied.
}
\label{fig:eloss_dist}
\end{figure}

\begin{figure}[]
\begin{center}
\subfloat[]{\label{fig:sudakov_safety_alphasweep}
\includegraphics[width=7cm]{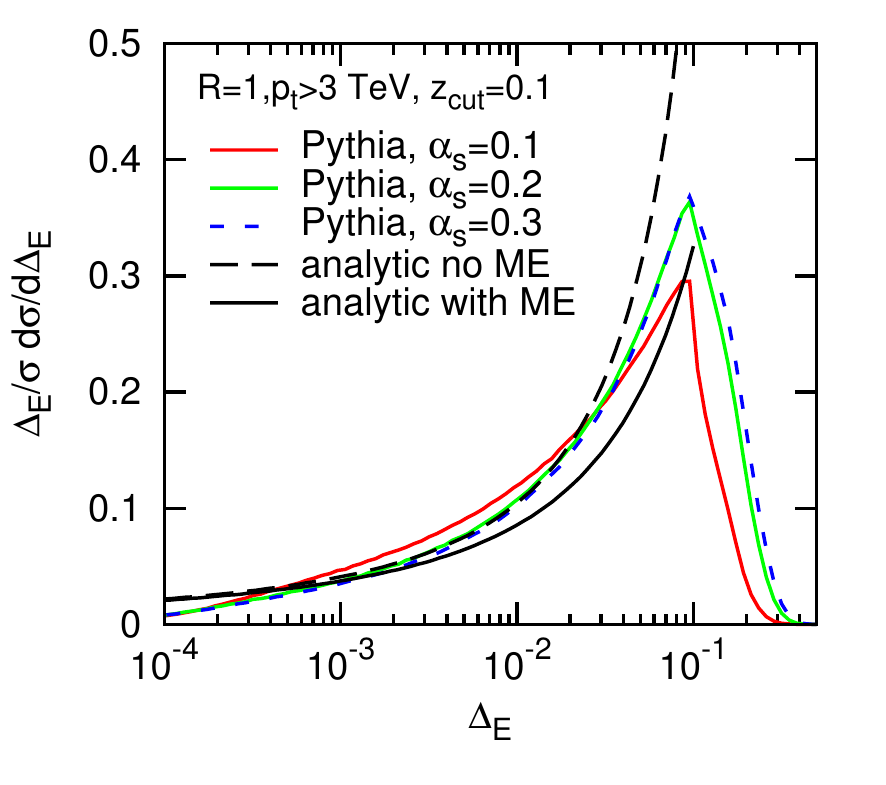}
}
$\quad$
\subfloat[]{\label{fig:sudakov_safety_ptsweep}
\includegraphics[width=7cm]{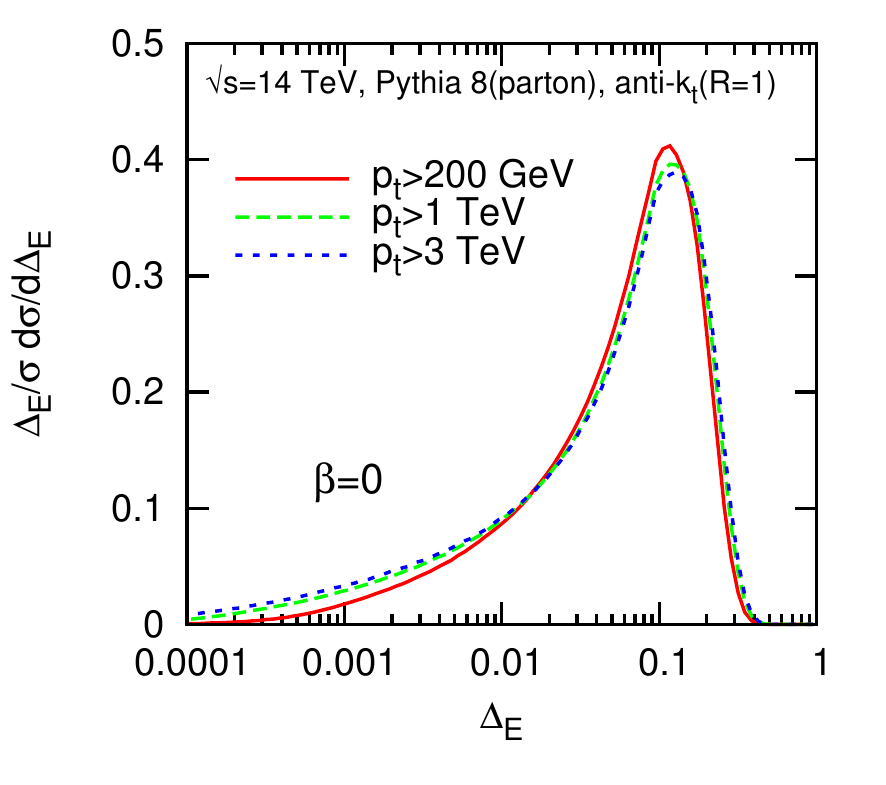}
}
\end{center}
\caption{The dependence of the $\beta=0$ energy drop distribution on $\alpha_s$.  On the left, we show \pythia{} results with fixed coupling compared to the fixed-coupling analytical prediction of \Eq{eq:energydropnoalpha}.  On the right, we show the $\Delta_E$ distribution with running coupling at different values of the jet's transverse momentum.  Both plots support the interpretation that the $\Delta_E$ distribution at $\beta=0$ is largely independent of $\alpha_s$.}
\label{fig:sudakov_safety}
\end{figure}

We conclude this section by comparing the fractional energy loss distribution between \pythia{8} to our MLL calculation, using the same jet samples as \Sec{sec:ang-MC}. The comparison is shown in \Fig{fig:eloss_dist}, with the Monte Carlo simulation on the left plot and our analytic calculation on the right.  On the analytic plots, the solid (dashed) curves represent the result with (without) the inclusion of the multiple emission contributions to $\Delta_E$. 

For $\beta>0$, there is good agreement between \pythia{} and our MLL analytics.  For the IRC unsafe (but Sudakov safe) limit $\beta=0$, the agreement is fair in the region $\Delta_E<\zcut$.  Note that $\beta = 0$ has a large contribution from multiple emissions, but the structure of the inverse Laplace transform enforces that the MLL result cannot extend beyond $\Delta_E = \zcut$.  In contrast, the \pythia{} distribution extends well beyond $\zcut$.  This effect from multiple hard emissions contributing to $\Delta_E > \zcut$ is not captured by our resummation. 

We can study the $\beta = 0$ limit in \pythia{} to see whether the analytic predictions of \Sec{sec:sdkv} are born out in Monte Carlo.  In \Fig{fig:sudakov_safety_alphasweep}, we show the $\Delta_E$ distribution for $\beta=0$ by artificially turning off the running coupling and setting the $\alpha_s$ value by hand.  As discussed in \Eq{eq:energydropnoalpha}, the fixed-coupling analytic resummation does not depend on $\as$. Indeed, we see that the Monte Carlo results are fairly independent of the $\alpha_s$ value, and the behavior is well described by the analytic calculation.  The same physical effect is seen in \Fig{fig:sudakov_safety_ptsweep}, where the running coupling is restored but the distribution is shown for different choices of the minimum transverse momentum of the jet, which in turn probes different values of $\alpha_s$.  We note that the curves differ very little from each other, suggesting that leading $\alpha_s$-independence of the $\beta = 0$ result is robust.

\section{Non-Perturbative Contributions}
\label{sec:NP}

\begin{figure}[p]
\begin{center}
\subfloat[]{\label{fig:hadr_C1}
\includegraphics[width=6cm]{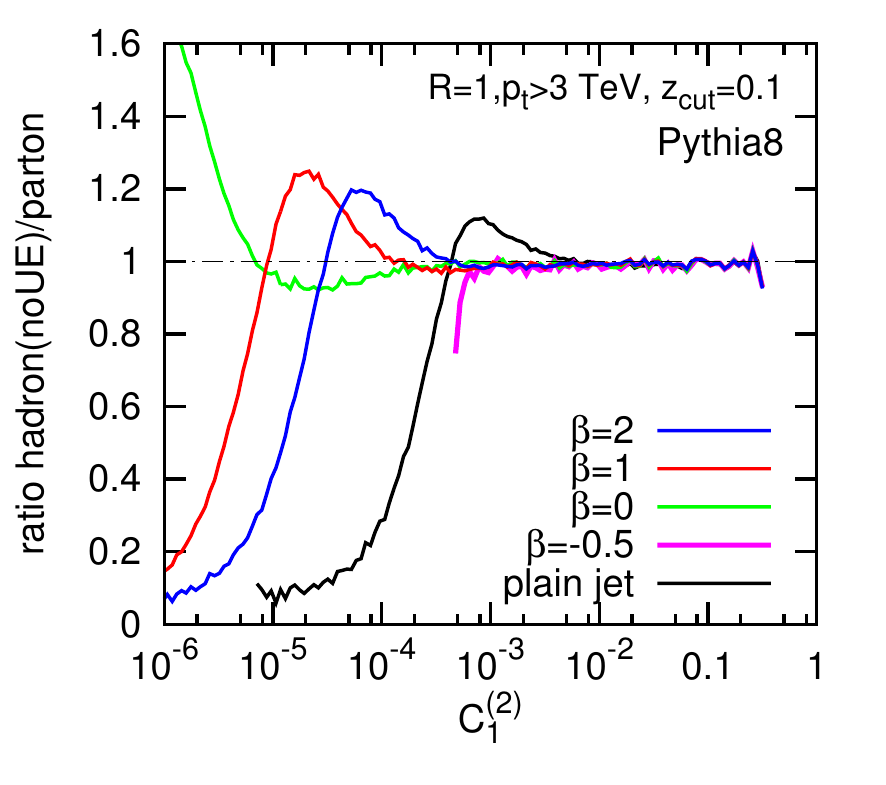}
}
$\quad$
\subfloat[]{\label{fig:UE_C1}
\includegraphics[width=6cm]{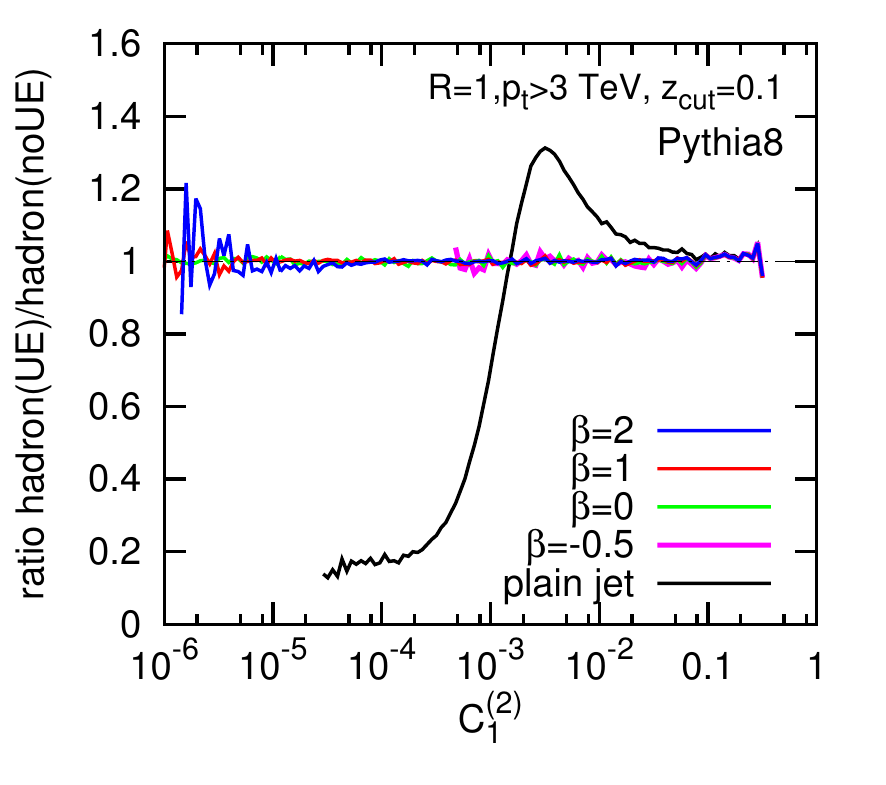}
}
~\\
\subfloat[]{\label{fig:hadr_Rg}
\includegraphics[width=6cm]{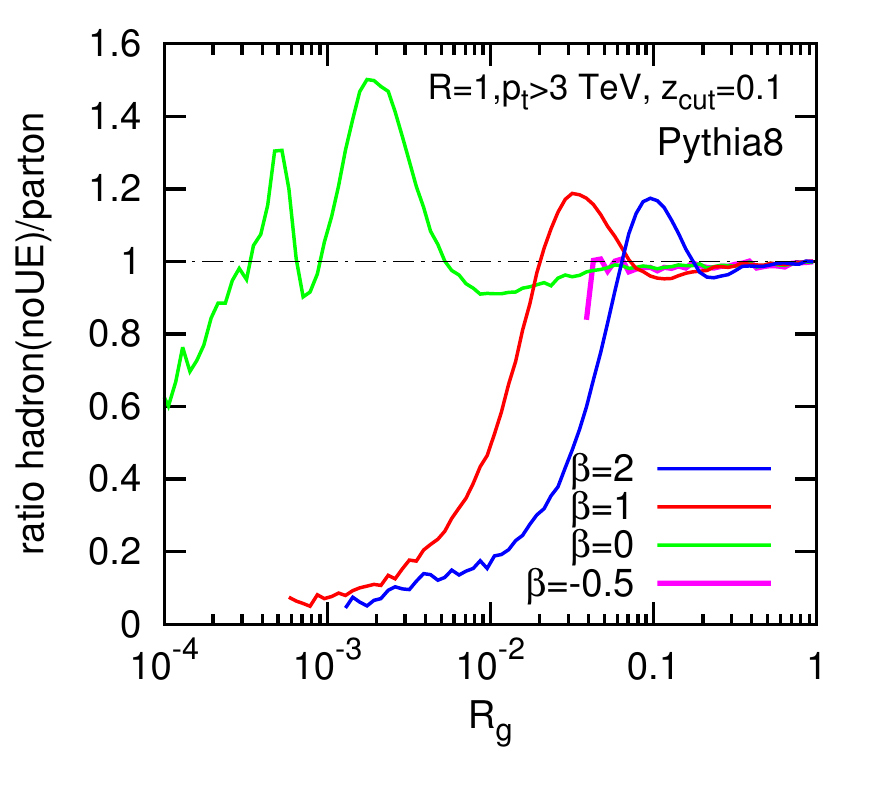}
}
$\quad$
\subfloat[]{\label{fig:UE_Rg}
\includegraphics[width=6cm]{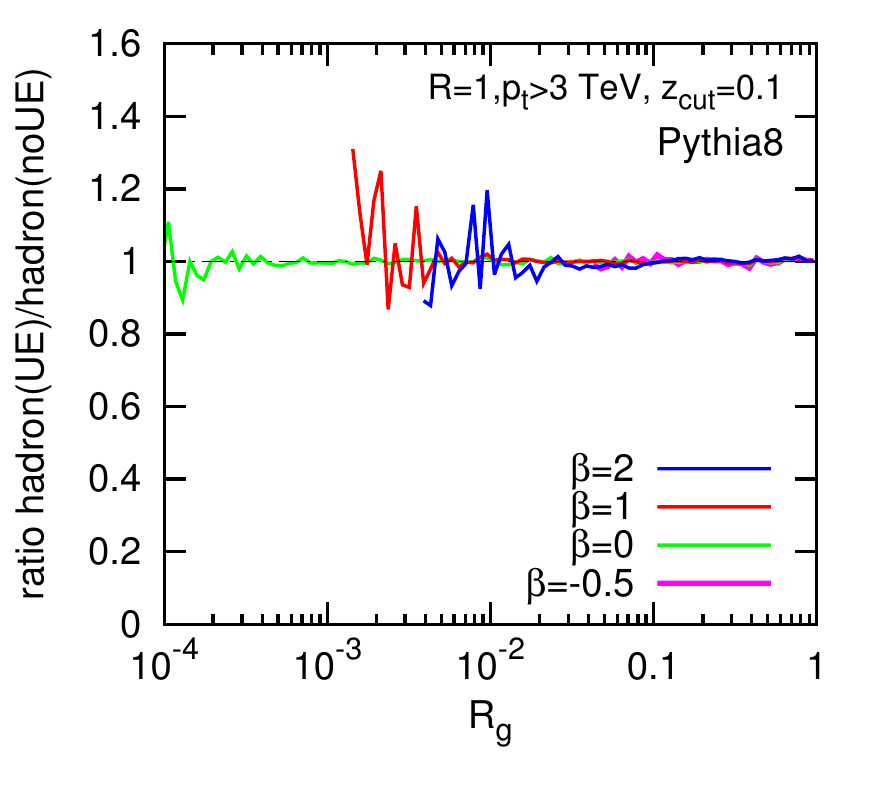}
}
~\\
\subfloat[]{\label{fig:hadr_DE}
\includegraphics[width=6cm]{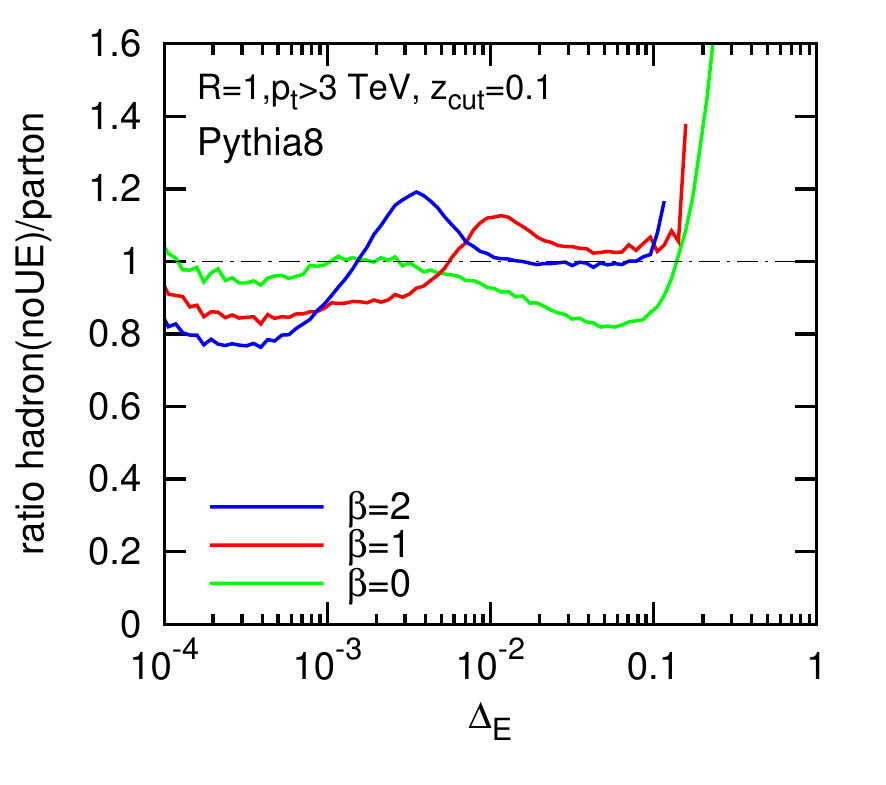}
}
$\quad$
\subfloat[]{\label{fig:UE_DE}
\includegraphics[width=6cm]{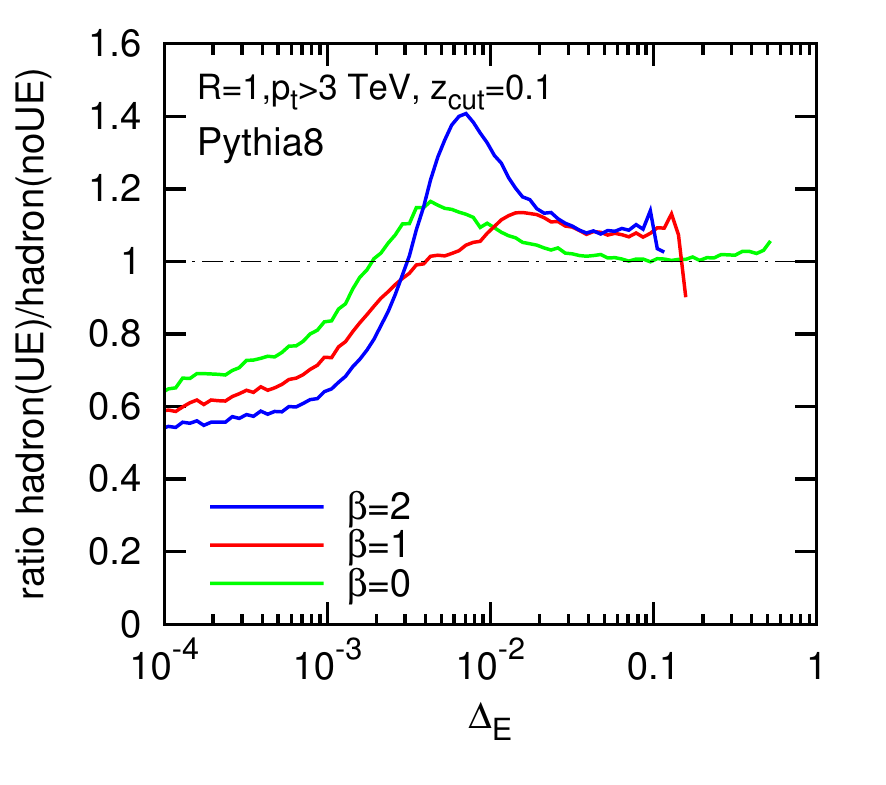}
}
\end{center}
\caption{Effect of non-perturbative corrections on $C_1^{(\alpha=2)}$ (top), $R_g$ (middle), and $\Delta_E$ (bottom).  The plots on the left show the ratio between hadron level and parton level predictions obtained with \pythia{8} (without UE). The plots on the right instead show the ratio of hadron-level results with and without UE.}
\label{fig:np_corrections}
\end{figure}

In all of the above analytic calculations, we only considered the distributions generated by perturbative partons.  In this section, we will do a brief Monte Carlo study to try to estimate the impact that non-perturbative effects from hadronization and UE can have on these distributions.

In \Fig{fig:np_corrections}, we show the effect of hadronization (left) and UE (right) for various observables considered in this paper.  In the case of hadronization, we plot the ratio between the hadronic and partonic distributions obtained from \pythia{8}. In the case of UE, we plot the ratio between the distributions with and without UE.  Apart from including non-perturbative effects, the details of the analysis are the same as for the previous Monte Carlo studies.

We start by considering $\ea$ for $\alpha=2$, i.e.\ similar to jet mass. The plot in \Fig{fig:hadr_C1} shows that soft drop declustering pushes the onset of hadronization corrections to smaller values of the observable compared to the ungroomed case (shown in black).  As shown in \Fig{fig:UE_C1}, soft drop has the remarkable ability to reduce the UE contribution to almost zero.

For the groomed jet radius distribution, the behavior of hadronization corrections in \Fig{fig:hadr_Rg} is qualitatively similar to those seen for $\ea$, with hadronization having a smaller effect for smaller values (and negative values) of $\beta$.  The UE event contribution to $R_g$ in \Fig{fig:UE_Rg} is also fairly small.

Finally, we show the effect of hadronization and UE corrections on the jet energy drop in \Figs{fig:hadr_DE}{fig:UE_DE}, respectively.  Unlike for the previous distributions, the hadronization corrections are largest for $\beta=0$, which is likely related to the issue of IRC unsafety.  For all values of $\beta$, the UE corrections are fairly large for $\Delta_E$.  That said, because $\Delta_E$ is defined in terms of both the groomed energy $E_g$ and the ungroomed energy $E_0$, it is hard to know whether these effects are caused mainly by $E_g$ or $E_0$.  We suspect that $E_g$ is rather robust to UE effects, and the dominant change is really from distortions of the reference $E_0$ value.

\section{Boosted $W$ Tagging with Soft Drop}
\label{sec:tagging}

\begin{figure}[p]
\begin{center}
\subfloat[]{\label{fig:ROC}
\includegraphics[width=7cm]{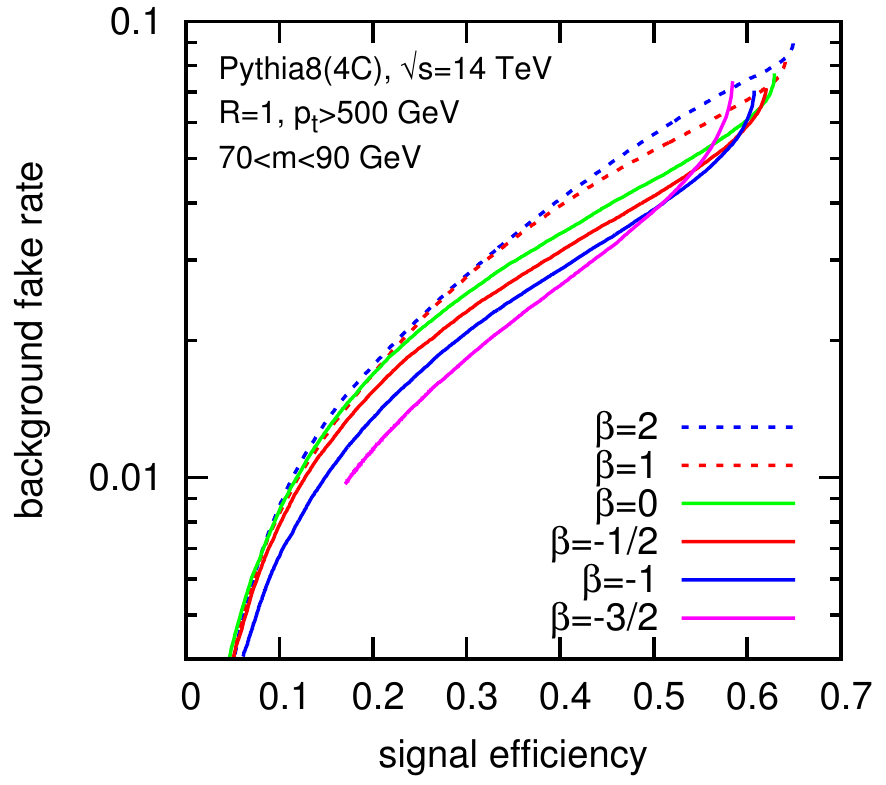}
}
$\quad$
\subfloat[]{\label{fig:zcut_values}
\includegraphics[width=7cm]{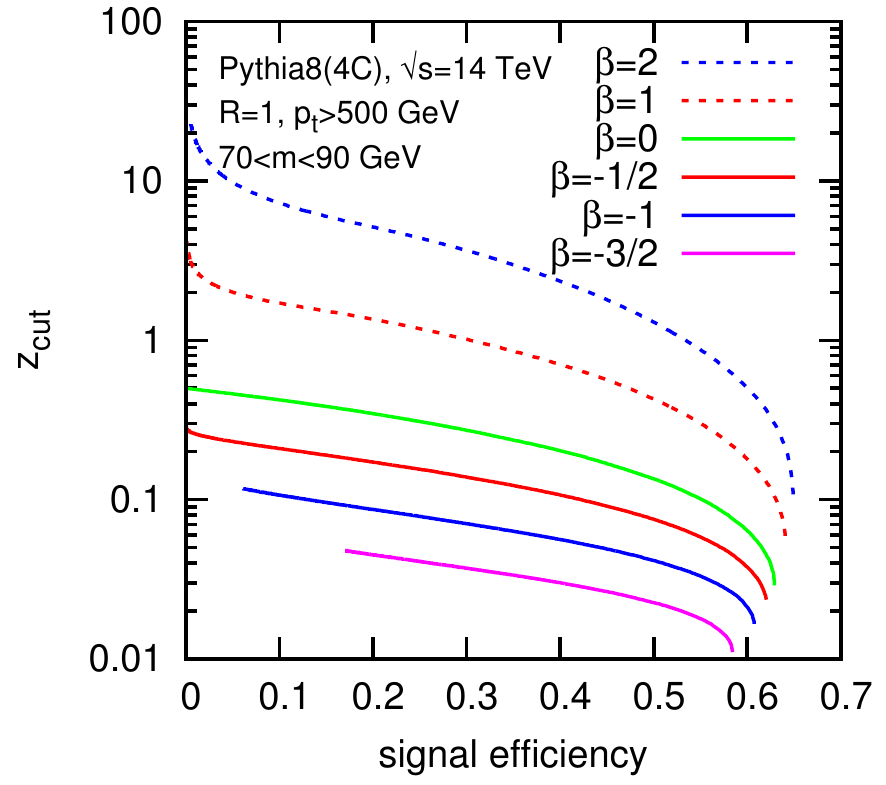}

}
~\\
\subfloat[]{\label{fig:signal_distr}
\includegraphics[width=7cm]{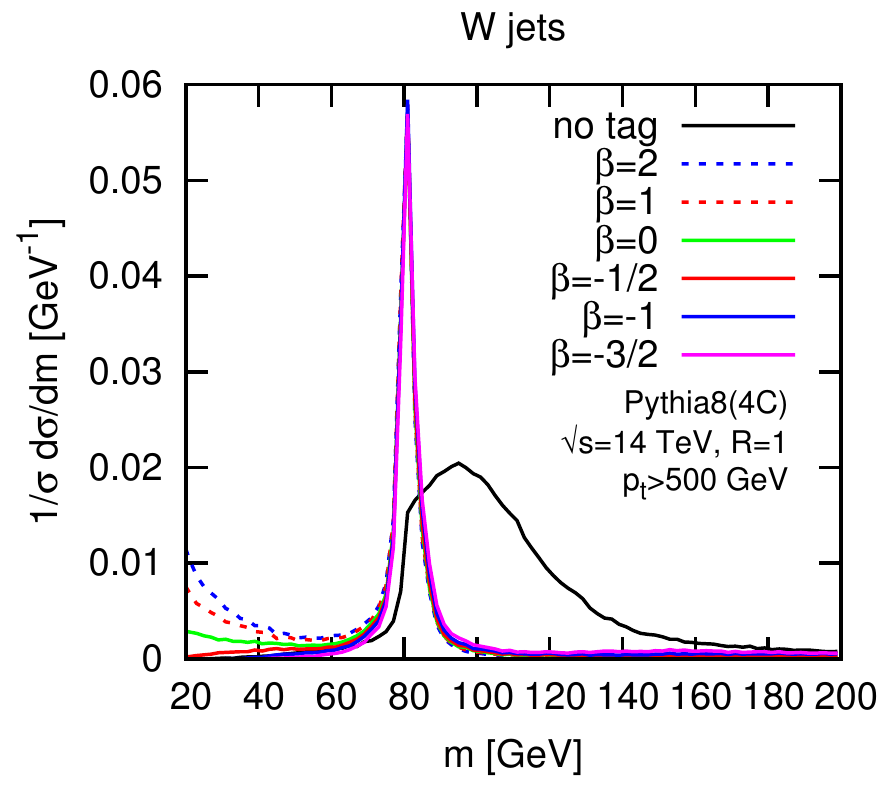}

}
$\quad$
\subfloat[]{\label{fig:background_distr}
\includegraphics[width=7cm]{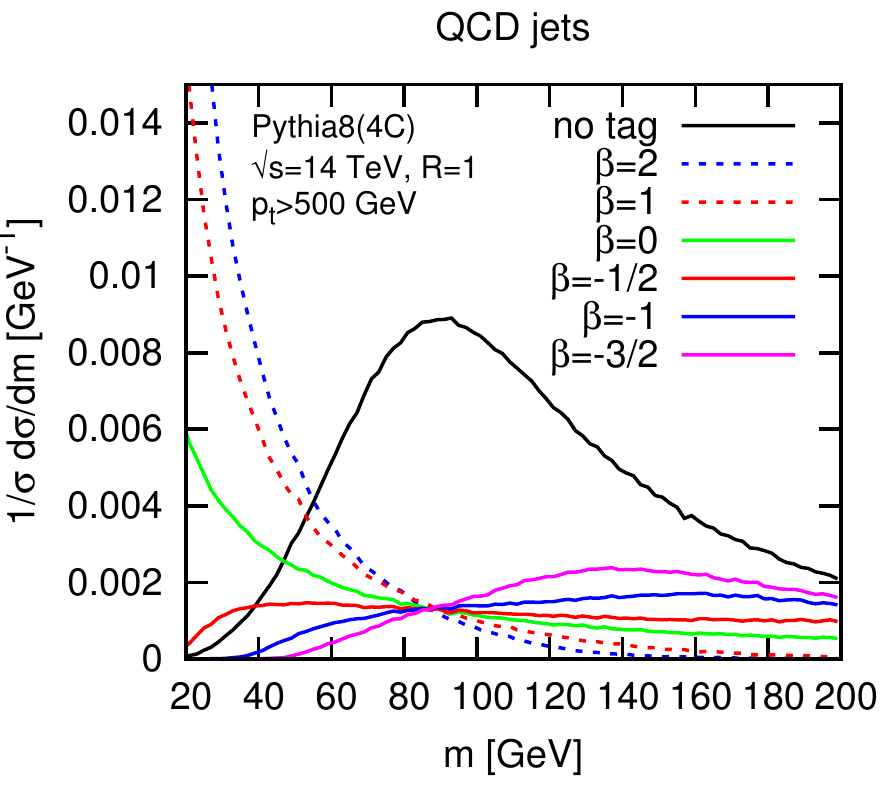}

}
\end{center}
\caption{Performance of soft drop as a boosted $W$ tagger.  Top left:  signal efficiency versus background mistag for jets with $p_T > 500~\text{GeV}$.  Each curve is obtained by fixing the value of $\beta$, sweeping the value of  $z_{\rm cut}$, and counting jets with groomed mass in the range $[70~\text{GeV}, 90~\text{GeV}]$. 
Top right: Values of $\zcut$ for as a function of the efficiency, for given $\beta$.
Bottom: mass distribution of signal (left) and background (right) jets before and after soft drop. 
For each curve, the value of $\beta$ is shown in the legend, while the value of $\zcut$ is the one that gives a 35\% signal efficiency.
}
\label{fig:tagging}
\end{figure}

\begin{figure}[]
\begin{center}
\includegraphics[width=7cm]{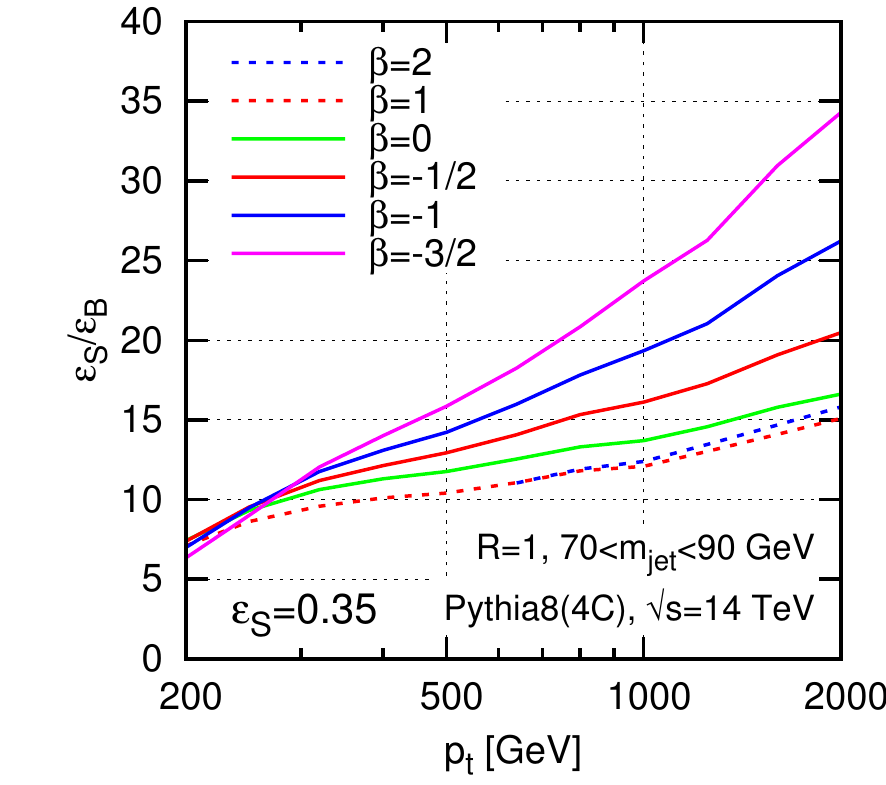}
\end{center}
\caption{Ratio of signal-to-background efficiency as a function of the minimum jet $p_{T}$ for fixed signal efficiency of 35\%.
}
\label{fig:efficiencies_vs_pt}
\end{figure}
Thus far, we have studied the analytic properties of soft drop declustering and argued that it can be a successful grooming technique for $\beta >0$.  For $\beta<0$, soft drop acts like a tagger which identifies jets with hard two-prong structures. Here, we investigate the performance of soft drop in tagging mode by doing a brief study of boosted $W$ tagging.

To have a source of fat $W$ and QCD jets, we generated $WW$ and dijet samples with \pythia{8} for 14 TeV proton-proton collisions, including all non-perturbative effects from tune 4C.  As in the previous Monte Carlo studies in this paper, we start from anti-$k_t$ jets with $R_0=1$, this time keeping only jets with $p_T\ge p_{T \min}$ and rapidity $|y|<4$.  These samples of $W$ (signal) and QCD (background) jets are then
groomed/tagged using soft drop with various values of $\beta$ and $z_{\rm cut}$, and we define the efficiency/mistag rates from the fraction of selected jets after soft drop with groomed masses in the $W$ window $[70~\text{GeV}, 90~\text{GeV}]$.

The results of this study are presented in \Figs{fig:tagging}{fig:efficiencies_vs_pt}.  In \Fig{fig:ROC}, we fix $p_{T \min} = 500~\text{GeV}$ and study the efficiency/mistag rates for fixed $\beta$, sweeping $z_{\rm cut}$.  The values of $\zcut$ found as a function of the efficiency are shown in \Fig{fig:zcut_values}.  As initially expected, negative values of $\beta$ (i.e.~tagging mode) tend to have a higher performance than positive values (i.e.~grooming mode).  Note also that the $\zcut$ values for $\beta < 0$ fall in the more reasonable range of $\zcut \lesssim 1$, whereas $\zcut \gtrsim 1$ is needed to obtain comparable performance for $\beta > 0$.

In \Figs{fig:signal_distr}{fig:background_distr}, we show the mass distributions of signal and background jets with $p_T>500$~GeV after soft-drop, for different values of $\beta$ and choosing the value of $\zcut$ that correspond to 35\% signal efficiency. Regarding the signal, all values of $\beta$ yield a nice narrow mass distributions around $m_W$.  Without soft drop, the background in this $p_T$ window happens to (accidentally) have a mass peak around $m_W$, but as desired, the soft-dropped background mass distributions are pushed away from the signal region.

Finally, in \Fig{fig:efficiencies_vs_pt}, we study the ratio of signal-to-background efficiency as a function of $p_{T \min}$ (at fixed 35\% signal efficiency).  Negative values of $\beta$ continue to have a higher performance, especially at large $p_T$, essentially because of a stronger Sudakov suppression of the background at fixed signal efficiency. The overall performance is comparable to other $W$ tagging methods, with percent-level mistag rates at 35\% efficiency.

The original mass-drop prescription from \Ref{BDRS} also involves a filtering step.  There, the filtering radius was taken as $\min(R_g/2, 0.3)$ with $R_g$ defined as in \Sec{sec:pileup}, and the three hardest subjets were kept. 
However, applying filtering on soft-dropped jets is not necessarily beneficial. For example, at large $p_T$ and for
$\beta<0$, the action of the soft drop is such that the background
peaks at a value of the mass larger than the $W$ mass. In this case, filtering would slightly
shift this peak to smaller masses, increasing the background
rates. On the other hand, we should add that in a situation with pileup, filtering
or some similar form of grooming might also be needed in order to
improve the resolution on the signal peak. We leave a detailed study of the interplay between pileup mitigation and boosted object tagging to future work.

\section{Conclusions}
\label{sec:conclude}

In this paper, we introduced the soft drop declustering procedure. Soft drop generalizes the mMDT procedure by incorporating an angular exponent $\beta$, and simplifies mMDT by removing the mass drop condition.   True to its name, the soft drop procedure drops wide-angle soft radiation from a jet, though for $\beta \le 0$ it can also drop collinear radiation.  To demonstrate the analytic behavior of soft drop declustering, we calculated three distributions to MLL accuracy (while also including multiple emissions):  the energy correlation functions $\C{1}{\alpha}$, the groomed jet radius $R_g$, and the jet energy drop $\Delta_E$.  Two particularly interesting analytic features are the smooth turn off of non-global logarithms for $\C{1}{\alpha}$ in the $\beta \to 0$ limit and the approximate $\alpha_s$-independence of the jet energy drop distribution for $\beta = 0$.  

Beyond our analytic calculations, we studied the performance of soft drop in two other contexts.  We used a Monte Carlo study to estimate the impact of non-perturbative effects, and found that soft drop reduces the impact of hadronization and UE corrections on $\C{1}{\alpha}$ compared to ungroomed case.  We also used a Monte Carlo study to demonstrate that soft drop with $\beta < 0$ can act as an effective tagger for boosted $W$ bosons.

One area for future study is the behavior of soft drop as a pileup mitigation tool.  We have seen that soft drop yields small values of $R_g$ and hence small jet areas, so one might expect that soft drop would have similar pileup performance to trimming \cite{trimming,ATLAS:2012jla,TheATLAScollaboration:2013ria,cms}.  Like trimming, soft drop declustering with $\beta > 0$ is an all-purpose grooming procedure, in the sense that the grooming procedure does not veto jets (unlike a tagger), and the groomed version of an (otherwise) IRC safe observable is still IRC safe.\footnote{This is in contrast to the $\beta = 0$ (mMDT) limit, which has to be run in tagging mode to obtain an IRC safe groomed jet energy distribution.  Of course, we have argued that $\beta = 0$ soft-dropped distributions are still Sudakov safe.}   Both trimming and soft drop have two parameters.  In the case of trimming, they are the energy fraction threshold $f_{\rm cut}$ and the subjet radius $R_{\rm sub}$.  In the case of soft drop, they are the soft drop threshold $\zcut$ and the angular exponent $\beta$.  The $f_{\rm cut}$ and $\zcut$ parameters play a similar role, since they control how aggressive the grooming procedure is and also define the transition points in, e.g., the $\ea$ distribution.  

However, there is a qualitative difference between $R_{\rm sub}$ and $\beta$ which is likely to be phenomenologically relevant.  At fixed values of the jet mass, harder jets become narrower jets.  The radius parameter $R_{\rm sub}$ sets a fixed angular scale, such that narrower jets are effectively groomed less (see the discussion in \Ref{taggersRES}).  In contrast, $\beta$ sets a scaling relation between energies and angles, such that the amount of grooming decreases only gradually as the jets become more narrow.  The extreme limit of $\beta = 0$ is where approximately the same fraction of energy is groomed away regardless of the initial jet energy (see \Eq{eq:energydropnoalpha}).  Thus, we expect that soft drop could potentially have better performance than trimming at higher energies and luminosities.  Of course, this assumes that detectors are able to resolve angular scales smaller than the typical $R_{\rm sub} \simeq 0.2$.

Finally, in a more speculative vein, one might wonder whether the soft drop procedure could be applied on an event-wide basis instead of jet-by-jet as considered here.  In the case of trimming, there is a suitable generalization \cite{Bertolini:2013iqa} such that the trimming criteria can be imposed without needing to first cluster an event into jets.  In the case of mass drop, there are ways to sew together different jet multiplicities to impose a kind of event-wide mass drop condition \cite{Gouzevitch:2013qca}.  If the soft drop condition in \Eq{eq:mainrelation} could be applied on an event-wide basis, this could help address many of the numerous complications associated with soft radiation and allow analyses to focus on the more tractable collinear physics of jets.

\begin{acknowledgments}
  We thank the organizers of the ESI Program on Jets and QFT and of
  Boost 2013 for stimulating workshops during which this project was
  started.  We acknowledge useful discussions with Mrinal Dasgupta,
  Gavin Salam and Jeff Tseng. S.M. would like to thank IPhT Saclay for hospitality during the course of this project.
    A.L. and J.T. are supported by the U.S. Department of
  Energy (DOE) under cooperative research agreement
  DE-FG02-05ER-41360. J.T. is also supported by the DOE Early Career
  research program DE-FG02-11ER-41741 and by a Sloan Research
  Fellowship from the Alfred P. Sloan Foundation. S.M. is supported by
  the UK Science \& Technology Facilities Council (STFC). G.S. is
  supported by the French Agence Nationale de la Recherche,
  under grant ANR-10-CEXC-009-01 and by the EU ITN grant LHCPhenoNet,
  PITN-GA-2010-264564.

\end{acknowledgments}

\appendix

\section{Details of Energy Correlation Calculation}
\label{app:angdist}

We present the details of the calculation of the soft-drop energy correlation function ($\ea$ with $\alpha>0$) to MLL accuracy.
Thus, we consider the independent emission of $n$ collinear gluons within a jet. For each splitting $m$, the scale of the (one-loop) coupling is chosen at the relative transverse momentum scale $\kappa_m = z_m \, \theta_m \, p_{T \text{jet}}$. This is sufficient to capture logarithmic accuracy we seek in this study (for a more detailed discussion, see the extensive literature on event-shape and jet-mass resummation, e.g.~\cite{Catani:1992ua,Dokshitzer:1998kz,caesar}).

The above undestanding translates into
\begin{align}\label{all-order-ang-start}
\Sigma(\ea)&= \sum_{n=0}^{\infty}\frac{1}{n!} \prod_{m=1}^n 
\int \frac{d \theta_m}{\theta_m} \int d z_m\,  p_i(z_m) \frac{ \as(\kappa_{m})}{\pi} \left[
\Theta\left(\zcut \left(\frac{\theta_m}{R_0} \right)^\beta -z_m \right) \right. \nonumber \\ & \quad ~ + \left.
\Theta\left(z_m- \zcut \left(\frac{\theta_m}{R_0} \right)^\beta \right)
\Theta \left(\ea -z_m \left(\frac{\theta_m}{R_0} \right)^\alpha\right)-1
 \right] \Theta\left(R_0-\theta_m \right)
  \nonumber \\ &
   =\sum_{n=0}^{\infty}\frac{(-1)^n}{n!} \prod_{m=1}^n 
\int \frac{d \theta_m}{\theta_m} \int  d z_m\, p_i(z_m) \frac{ \as(\kappa_{m})}{\pi}  \left[
\Theta \left(z_m\left(\frac{\theta_m}{R_0} \right)^\alpha-\ea \right) \right. \nonumber \\ & \quad ~ \left. \times
\Theta\left(z_m-\zcut \left(\frac{\theta_m}{R_0} \right)^\beta\right) \right] \Theta\left(R_0-\theta_m \right).
\end{align}
To MLL accuracy, the resummed result is then
\begin{equation}
\Sigma(\ea) = e^{-R(\ea)} \ ,
\end{equation}
where the radiator is given by the integral of the one-loop contribution over the allowed phase-space:
\begin{equation}\label{rad-ang-end-app}
R(\ea)= \frac{1}{\alpha}\int_{\ea}^1 \frac{d c}{c} 
\int_{\max \left(c,\, {\zcut}^\frac{\alpha}{\alpha+\beta} {c}^\frac{\beta}{\alpha+\beta} \right) }^1 d z\, p_i(z)
\frac{\as\left(\kappa  \right)}{\pi}
  \ .
\end{equation}
The reduced splitting functions $p_i$, with $i=q,g$ are given by
\begin{subequations}
  \label{eq:reduced-splitting}
  \begin{align}
  p_{q}(z)& = C_F \frac{1+(1-z)^2}{z}\,,\\
 p_{g} (z)&= C_A\left[ 2\frac{1-z}{z} + z(1-z)+ \frac{T_R n_f}{C_A} (z^2 + (1-z)^2)\right]\,.
  \end{align}
\end{subequations}

Note that for small enough values of energy fractions $z$ and angular distances $\theta$, the argument of the coupling in \Eq{rad-ang-end-app} can approach the non-perturbative region.  Thus, we introduce a prescription in order to evaluate the integrals down to these low scales. We decide to freeze the coupling below a non-perturbative scale $\mu_\text{NP}$:
\begin{equation}\label{eq:coupling-freezing}
\as (\kappa)= \as^\text{1-loop}(\kappa)\Theta\left (\kappa-\mu_\text{NP}\right)+\as^\text{1-loop}(\mu_\text{NP})\Theta\left (\mu_\text{NP}-\kappa\right)\,,
\end{equation}
where $\as^\text{1-loop}(\kappa)$ is the usual one-loop expression for
the strong coupling, i.e.\ its running is evaluated with $\beta_0$ only:
\begin{equation}\label{eq:coupling-1loop}
\as^\text{1-loop}(\kappa)=\frac{\as(Q)}{1+2 \as(Q) \beta_0 \log \frac{\kappa}{Q}} \ .
\end{equation}
Our results are expressed in terms of $\as=\as(R_0\, p_T)$ and
we use $\as(m_Z)=0.12$, $n_f=5$, and $\mu_\text{NP}=1$~GeV
throughout this paper.

In the $(\log \frac{1}{z},\log \frac{R_0}{\theta})$ plane of \Fig{fig:groomedregions-C1}, the boundary between perturbative and non-perturbative regions is given by
\begin{equation}
\label{eq:freezingline}
\kappa = \mu_{\rm NP}  \quad \Rightarrow \quad \log \frac{1}{z} = \log\frac{1}{\tilde{\mu}}-\log \frac{R_0}{\theta},
\end{equation}
where we have introduced $\tilde{\mu}=\frac{\mu_{\rm NP}}{p_T R_0}$.
Thus, according to \Eq{eq:coupling-freezing}, below this straight line, the one-loop running coupling is evaluated at the relative transverse momentum $\kappa$, while above this line it is frozen at $\mu_{\rm NP}$.

The explicit form for the non-perturbative region of result depends on the relation between the slope of lines of constant $\ea$, which is controlled by $\alpha$, and the slope of the boundary between perturbative and non-perturbative regime, given by \Eq{eq:freezingline}.  We also assume $ \beta\ge0$ for simplicity and discuss the case $\beta<0$ in the end.

For $\alpha>1$ we find
\begin{align}\label{full-rad-ang-beta-pos-alpha-gt-1}
R(\ea) 
& \overset{\ea>\zcut}{\wideeq{2.0cm}}  \frac{C_i}{2\pi \alpha_s \beta_0^2} \bigg[
    \frac{\xlog(1-\lambda)}{\alpha-1}
     -\frac{\alpha \xlog(1-\frac{1}{\alpha}\lambda)}{\alpha-1}  
     -2 \alpha_s \beta_0 B_i \log(1-\frac{1}{\alpha}\lambda)\bigg]\\
& \overset{\zcut^\frac{1-\alpha}{1+\beta}\tilde{\mu}^\frac{\alpha+\beta}{1+\beta}< \, \ea<\zcut}{\wideeq{2.0cm}}  \frac{C_i}{2\pi \alpha_s \beta_0^2} \bigg[
 -\frac{\xlog(1-\lambda_c)}{1+\beta}
     -\frac{\alpha \xlog(1-\frac{1}{\alpha}\lambda)}{\alpha-1}  
     -2 \alpha_s \beta_0 B_i \log(1-\frac{1}{\alpha}\lambda)\notag \\& \phantom{\wideeq{2.0cm}}+
     \frac{\alpha+\beta}{(\alpha-1)(1+\beta)} \xlog\left(1-\frac{1+\beta}{\alpha+\beta}\lambda-\frac{\alpha-1}{\alpha+\beta}\lambda_c\right)
     \bigg] \label{eq:first1}
\\
& \overset{\tilde{\mu}^\alpha<\ea<\zcut^\frac{1-\alpha}{1+\beta}\tilde{\mu}^\frac{\alpha+\beta}{1+\beta}}{\wideeq{2.0cm}} \frac{C_i}{2\pi \alpha_s \beta_0^2} \bigg[ 
-\frac{\xlog(1-\lambda_c)}{1+\beta}
     -\frac{\alpha \xlog(1-\frac{1}{\alpha}\lambda)}{\alpha-1}  
     -2 \alpha_s \beta_0 B_i \log(1-\frac{1}{\alpha}\lambda)\notag \\& \phantom{\wideeq{2.0cm}}-\frac{1+\log\left(1-\lambda_\mu \right)}{(\alpha-1)(1+\beta)} \left((\alpha-1) \lambda_c+(1+\beta) \lambda-(\alpha+\beta) \lambda_\mu) \right)\notag \\& \phantom{\wideeq{2.0cm}}+
     \frac{\alpha+\beta}{(\alpha-1)(1+\beta)} \xlog\left(1-\lambda_\mu \right)
 \bigg] +
  \frac{C_i \as(\mu_\text{NP})}{\pi} \, \mathcal{F}_1(L)
\\
& \overset{\ea<\tilde{\mu}^\alpha}{\wideeq{2.0cm}} \frac{C_i}{2\pi \alpha_s \beta_0^2} \bigg[ 
-\frac{\xlog(1-\lambda_c)}{1+\beta}
     -\frac{\beta \xlog(1-\lambda_\mu)}{1+\beta}  
     -2 \alpha_s \beta_0 B_i \log(1-\lambda_\mu)\notag \\& \phantom{\wideeq{2.0cm}}-\frac{1+\log\left(1-\lambda_\mu \right)}{(1+\beta)} \left( \lambda_c+\beta \lambda_\mu \right)
 \bigg]
 + \frac{C_i \as(\mu_\text{NP})}{\pi} \, \bigg[\mathcal{F}_1(\alpha \, L_\mu)
  \notag \\& \phantom{\wideeq{2.0cm}}
  + \left(\frac{L}{\alpha}- L_\mu\right)\left(\frac{2\alpha}{\alpha+\beta} L_c+2 B_i +\frac{\beta}{\alpha+\beta}(L+\alpha L_\mu) \right)
 \bigg],
 \end{align}
while for $\alpha<1$ we have\footnote{For definiteness we consider the case $\zcut>\tilde{\mu}^\alpha$.}
 \begin{align}\label{full-rad-ang-beta-pos-alpha-lt-1}
R(\ea) 
& \overset{\ea>\zcut}{\wideeq{2.0cm}}  \frac{C_i}{2\pi \alpha_s \beta_0^2} \bigg[
    \frac{\xlog(1-\lambda)}{\alpha-1}
     -\frac{\alpha \xlog(1-\frac{1}{\alpha}\lambda)}{\alpha-1}  
     -2 \alpha_s \beta_0 B_i \log(1-\frac{1}{\alpha}\lambda)\bigg]\\
& \overset{\tilde{\mu}^\alpha< \, \ea<\zcut}{\wideeq{2.0cm}}  \frac{C_i}{2\pi \alpha_s \beta_0^2} \bigg[
 -\frac{\xlog(1-\lambda_c)}{1+\beta}
     -\frac{\alpha \xlog(1-\frac{1}{\alpha}\lambda)}{\alpha-1}  
     -2 \alpha_s \beta_0 B_i \log(1-\frac{1}{\alpha}\lambda)\notag \\& \phantom{\wideeq{2.0cm}}+
     \frac{\alpha+\beta}{(\alpha-1)(1+\beta)} \xlog\left(1-\frac{1+\beta}{\alpha+\beta}\lambda-\frac{\alpha-1}{\alpha+\beta}\lambda_c\right)
     \bigg]
\\
& \overset{\zcut^\frac{1-\alpha}{1+\beta}\tilde{\mu}^\frac{\alpha+\beta}{1+\beta}<\ea<\tilde{\mu}^\alpha}{\wideeq{2.0cm}} \frac{C_i}{2\pi \alpha_s \beta_0^2} \bigg[ 
-\frac{\xlog(1-\lambda_c)}{1+\beta}
     -\frac{\alpha \xlog(1-\lambda_\mu)}{\alpha-1}  +
   \frac{\lambda-\alpha \lambda_\mu}{\alpha-1}\left(1+\log(1-\lambda_\mu) \right)
    \notag \\& \phantom{\wideeq{2.0cm}} -2 \alpha_s \beta_0 B_i \log(1-\lambda_\mu)+\frac{\alpha+\beta}{(\alpha-1)(1+\beta)} \xlog\left(1-\frac{1+\beta}{\alpha+\beta}\lambda-\frac{\alpha-1}{\alpha+\beta}\lambda_c\right)
 \bigg] \notag \\& \phantom{\wideeq{2.0cm}}+
  \frac{C_i \as(\mu_\text{NP})}{\pi} \,  \left(\frac{L}{\alpha}- L_\mu\right)\left(\frac{L-\alpha L_\mu}{1-\alpha}+2B_i\right)
  \\
& \overset{\ea<\zcut^\frac{1-\alpha}{1+\beta}\tilde{\mu}^\frac{\alpha+\beta}{1+\beta}}{\wideeq{2.0cm}} \frac{C_i}{2\pi \alpha_s \beta_0^2} \bigg[ 
-\frac{\xlog(1-\lambda_c)}{1+\beta}
     -\frac{\beta \xlog(1-\lambda_\mu)}{1+\beta}  -
 \left(\frac{\lambda_c+\beta \lambda_\mu}{1+\beta} \right)
   \notag \\& \phantom{\wideeq{2.0cm}} \times
 \left(1+\log(1-\lambda_\mu) \right)-2 \alpha_s \beta_0 B_i \log(1-\lambda_\mu)
 \bigg] \notag \\& \phantom{\wideeq{2.0cm}}
 + \frac{C_i \as(\mu_\text{NP})}{\pi} \, \bigg[\mathcal{F}_2(L)
 +\frac{(1-\alpha) (\beta L_\mu+L_c) (2 (1+\beta) B_i+\beta L_\mu+L_c)}{\alpha (1+\beta)^2}
  \bigg].
 \end{align}
In the above expressions, we have introduced $\xlog(x) = x\log x$ and
\begin{align}
\mathcal{F}_1(L)&= \frac{((1+\beta) L-(\alpha+\beta)L_\mu+(\alpha-1) L_c)^2}{(\alpha-1) (1+\beta) (\alpha+\beta)}, \\
\mathcal{F}_2(L)&=  \frac{(1+\beta) L-(\alpha+\beta)L_\mu+(\alpha-1) L_c}{\alpha (1+\beta)^2 (\alpha+\beta)} 
\\ \nonumber &\times
\left( \beta \left(\alpha+\beta\right) L_\mu +2 B_i(1+\beta) (\alpha+\beta)+L_c ( 2 \alpha+
   \alpha \beta+\beta)+\beta \left(1+\beta\right) L \right).
\end{align}
Here, $C_i$ is the color of the jet appropriate for quarks ($C_q = C_F$) or gluons ($C_g = C_A$).  $B_i$ describes the contribution to the cross section from collinear logarithms: $B_q = -3/4$ for quark jets and $B_g = -\frac{11}{12}+\frac{n_f}{6C_A} $ for gluon jets, where  $n_f$ is the number of active quark flavors.
We have also introduced
\begin{align} \label{log-refs}
&L=\log(1/\ea) && L_c=\log(1/\zcut) && L_\mu=\log(1/\mut),\\
&\lambda=2\alpha_s \beta_0 L && \lambda_c=2\alpha_s \beta_0 L_c && \lambda_\mu=2\alpha_s \beta_0 L_\mu.
\end{align}
Moreover, it can be easily checked that the limit $\alpha\to 1$ is perfectly safe because the two non-perturbative transition points coincide and therefore one non-perturbative region disappears. 

We note that expression for the running coupling with non-perturbative freezing \Eq{eq:coupling-freezing} has a discontinuous first derivative at $\kappa=\mu_\text{NP}$. To our logarithmic accuracy, this behavior is reflected into a discontinuity of the second derivative of the radiator at $\ea=\tilde{\mu}^\alpha$, which in turns causes a kink in the spectrum.
The difference between right- and left- second derivatives of the radiator at $\ea=\tilde{\mu}^\alpha$ is
\begin{equation}\label{eq:NP-disc}
c=\left( \frac{\as(\mu_\text{NP})}{\pi}\right)^2 \frac{4 \pi \beta_0 C_F B_i}{\alpha^2},
\end{equation}
which is a contribution beyond the accuracy of our calculation.
This effect is bigger for smaller $\alpha$, and for the case $\alpha=0.5$ in \Fig{fig:C1alpha05analytic}, the non-perturbative transition point $\ea=\tilde{\mu}^\alpha$ occurs in the vicinity of the Sudakov peak. Clearly, this is an artefact of our choice of an abrupt freezing of the coupling in the non-perturbative region. One could imagine to alter \Eq{eq:coupling-freezing} in such a way that it smoothly interpolates between running and fixed coupling (as done, for example, with scale profiling~\cite{Abbate:2010xh}). Alternatively, one could add an appropriate (subleading) term to the radiator in the region $\ea<\tilde{\mu}^\alpha$. However, we have decided not to introduce an ad-hoc prescription and, in this paper, we present results obtained from our MLL calculations (plus eventually multiple-emission effects), with the freezing of coupling previously discussed.

As already mentioned, the results for the resummed exponent have been obtained assuming $\beta>0$. It is clear from the expressions above that the $\beta \to 0 $ limit is perfectly safe.  Indeed for $\beta=0$ the result considerably simplifies and one obtains the mMDT single-logarithmic distribution. Moreover, the same results also hold for the $\beta<0$ case, provided that $\ea > \zcut^{\alpha/|\beta|}$, which is the minimum allowed value for the energy correlation function. For $\ea < \zcut^{\alpha\|\beta|}$, the radiator freezes at $R\left(\zcut^{\alpha/|\beta|}\right)$ and consequently the differential distribution vanishes.

Finally, the above results are also sufficient to compute the multiple-emission contributions described in Sec.~\ref{sec:pme}, which simply involve the derivative, as defined in \Eq{finite-derivative}, of the radiator functions derived in this appendix.

\section{Details of Jet Radius Calculation}
\label{app:rdist}

Here, we present the details of the calculation of the cumulative cross section of the jet radius after soft drop declustering. Because we are interested in the behavior of soft drop as a grooming procedure, we only consider $\beta>0$. As presented in \Sec{sec:pileup}, the cumulative resummed cross section can be computed from the sum over emissions as
\begin{align}\label{eq:rg_cum}
\Sigma^\text{radius}(R_g) &= \sum_{n=1}^\infty \frac{1}{n!} \prod_{m=1}^n\left[ \int_{R_g}^{R_0} \frac{d\theta_m}{\theta_m} \int_0^1 dz_m \, p_i(z_m)\, \frac{\alpha_s(\kappa_{m})}{\pi} \, \Theta\left(  z_\text{cut}\frac{\theta_m^\beta}{R_0^\beta}-z_m \right) \right] 
\nonumber \\ 
& \qquad\qquad \times 
e^{-\int_{R_g}^{R_0} \frac{d\theta}{\theta}\int_0^1 dz\, p_i(z)\, \frac{\alpha_s(\kappa)}{\pi}}\nonumber \\
&= e^{-R_1(R_g)} \ ,
\end{align}
where the exponent $R_1(R_g)$ is given by
\begin{equation}\label{eq:rexp}
R_1(R_g)=\int_{R_g}^{R_0} \frac{d\theta}{\theta}\int_0^1 dz\, p_i(z) \, \frac{\alpha_s(\kappa)}{\pi}\Theta\left( z- z_\text{cut}\frac{\theta^\beta}{R_0^\beta} \right) \ .
\end{equation}

The evaluation of the integrals proceed analogously to the case of the energy correlation case described in detail in \App{app:angdist}. 
In this case, the radiator is found to be
\begin{align} \label{radiator-rad}
R_1(R_g) 
& \overset{R_g>R_g^{(0)}}{\wideeq{2.0cm}}  \frac{C_i}{2\pi \alpha_s \beta_0^2} \bigg[
     -\xlog(1-\lambda_g)
     -\frac{\xlog(1-\lambda_c)}{1+\beta}
     +\frac{1}{1+\beta}\xlog(1-\lambda_c-(1+\beta)\lambda_g)\notag\\
     & \phantom{\wideeq{2.0cm}\frac{C_i}{2\pi \alpha_s \beta_0^2}\;}
     -2 \alpha_s \beta_0 B_i \log(1-\lambda_g)\bigg]\\
& \overset{\mu R_0<R_g<R_g^{(0)}}{\wideeq{2.0cm}}  \frac{C_i}{2\pi \alpha_s \beta_0^2} \bigg[
     -\xlog(1-\lambda_g)
     -\frac{\xlog(1-\lambda_c)}{1+\beta}
     +\frac{1-\lambda_c-(1+\beta)\lambda_g}{1+\beta} \log(1-\lambda_\mu)\notag\\
     & \phantom{\wideeq{2.0cm}\frac{C_i}{2\pi \alpha_s \beta_0^2}\;}
     +\frac{\lambda_\mu-\lambda_c-(1+\beta)\lambda_g}{1+\beta} -2 \alpha_s \beta_0 B_i \log(1-\lambda_g) \bigg]\notag\\
     & \phantom{\wideeq{2.0cm}}+\frac{C_i \alpha_s(\mu_{\rm NP})}{\pi} \frac{1}{1+\beta} \left[(1+\beta)L_g+L_c-L_\mu\right]^2\\
& \overset{R_g<\mu R_0}{\wideeq{2.0cm}} \frac{C_i}{2\pi \alpha_s \beta_0^2} \bigg[
     -\frac{\xlog(1-\lambda_c)}{1+\beta}
     -\frac{\lambda_c+\beta}{1+\beta} \log(1-\lambda_\mu)
     -\frac{\lambda_c+\beta \lambda_\mu}{1+\beta}\notag\\
     & \phantom{\wideeq{2.0cm}\frac{C_i}{2\pi \alpha_s \beta_0^2}\;}
     -2\alpha_s \beta_0 B_i\log(1-\lambda_\mu)\bigg]\notag\\
     & \phantom{\wideeq{2.0cm}}+ \frac{C_i\alpha_s(\mu_{\rm NP})}{\pi} \left[\frac{1}{1+\beta} (\beta L_\mu+L_c)^2+(L_g-L_\mu) (\beta L_g+\beta L_\mu+2 L_c+2 B_i)\right]
\end{align}
with $R_g^{(0)}=R_0 (\mut/\zcut)^{1/(1+\beta)}$ and $L_g=\log(R_0/R_g)$,
$\lambda_g=2\alpha_s \beta_0 L_g$.

\section{Details of Energy Drop Calculation}\label{app:edist}

Here, we present the details of the calculation of the cumulative cross section for the fractional energy drop from soft drop declustering.
Because we are interested in behavior of soft drop as a grooming procedure, we only consider $\beta>0$.
As opposed to the calculations previously described, for the energy drop distribution we also consider the effect of multiple emissions.
The resummed cumulative distribution is most easily written at fixed groomed jet radius $R_g$. The resulting expression is then integrated over all possible values of $R_g$:
\begin{equation}\label{eq:de_cum}
\Sigma^\text{energy-drop}(\Delta_E) = \int_0^{R_0} dR_g \, \frac{d \Sigma^\text{radius}(R_g) }{d R_g}\int \frac{d\nu}{2\pi i \nu}e^{\nu \Delta_E}  e^{-R_2\left(R_g,\nu^{-1}\right)} \ ,
\end{equation}
and $R_2(R_g,\nu)$ is the radiator function:
\begin{equation}\label{eq:rad_e}
R_2\left(R_g,\nu^{-1}\right)=\int_{R_g}^{R_0}\frac{d\theta}{\theta} \int_0^1 dz\, p_i(z) \frac{\alpha_s(\kappa)}{\pi}\Theta\left( z_\text{cut}\frac{\theta^\beta}{R_0^\beta}-z  \right)\left( 1- e^{-\nu z}  \right) \ ,
\end{equation}

In order to capture the single-logarithmic terms in \Eq{eq:rad_e} arising from multiple emissions we can make the following simplification~\cite{Catani:1992ua,Dokshitzer:1998kz,caesar}
\begin{equation}
{R}_2\left(R_g,\nu^{-1}\right)\simeq \bar{R}_2\left(R_g,\nu^{-1}\right) + \gamma_E \bar{R}'_2\left(R_g,\nu^{-1}\right) \ ,
\end{equation}
where
\begin{equation}
\bar{R}_2\left(R_g,\nu^{-1}\right)=2\int_{R_g}^{R_0}\frac{d\theta}{\theta} \int_{\nu^{-1}}^1 \frac{dz}{z}\, \frac{\alpha_s(\kappa)}{\pi}\Theta\left( z_\text{cut}\frac{\theta^\beta}{R_0^\beta}-z  \right) \ ,
\end{equation}
$\gamma_E$ is the Euler-Mascheroni constant and $\bar{R}_2'\left(R_g,\nu^{-1}\right)$ is the logarithmic derivative of $\bar{R}_2$ with respect to $\nu$. Moreover, note that we were able to drop the finite contributions to the splitting function $p_i(z)$ because 
for small values of $\zcut$, there are no logarithms from hard collinear emission. 

The inverse Laplace transform in \Eq{eq:de_cum} can be done to single logarithmic accuracy in $\nu$, also, by expanding $\nu$ about a fixed value $\nu_0$.  Doing this, the inverse Laplace transform becomes
\begin{equation}
\int \frac{d\nu}{2\pi i \nu}e^{\nu \Delta_E}  e^{-R_2\left(R_g,\nu^{-1}\right)} = \frac{\left(  \nu_0 \Delta_E \right)^{-\bar{R}'_2\left(R_g,\nu_0^{-1}\right)}}{\Gamma \left(1+\bar{R}'_2\left(R_g,\nu_0^{-1}\right)\right)}e^{-\bar{R}_2\left(R_g,\nu_0^{-1}\right) - \gamma_E \bar{R}_2'\left(R_g,\nu_0^{-1}\right)} \ .
\end{equation}
To minimize the logarithms, we choose $\nu_0=\Delta_E^{-1}$ and so the cumulative distribution of the groomed jet energy drop becomes
\begin{equation}
\Sigma^\text{energy-drop}(\Delta_E) = \int_0^{R_0} dR_g \, \frac{d \Sigma^\text{radius}(R_g)}{d R_g} \frac{e^{- \gamma_E \bar{R}'_2(R_g,\Delta_E)}}{\Gamma (1+\bar{R}_2'(R_g,\Delta_E))}e^{-\bar{R}_2(R_g,\Delta_E)} \ ,
\end{equation}

The evaluation of the integrals with running coupling proceeds in the same way as discussed for the energy correlation and groomed-jet radius distributions. 
We first obtain the energy drop cumulative distribution at fixed $R_g$ and then numerically integrate of $R_g$.
The radiator $\bar{R}_2$ is better described in three regions of $R_g$. First, for $R_g >R_0 (\mut/\zcut)^{1/(1+\beta)}$, we find
\begin{align}
\bar{R}_2(\Delta_E)
 & \overset{\Delta_E > \zcut (R_g/R_0)^\beta}{\wideeq{2.5cm}} \frac{C_i}{2\pi \alpha_s \beta_0^2} \bigg[
     \frac{\xlog(1-\lambda_c)}{1+\beta}
     -\xlog(1-\lambda_E)
     +\frac{\beta}{1+\beta} \xlog\left(1+\frac{\lambda_c-(1+\beta) \lambda_E}{\beta}\right) \bigg]\\
 & \overset{\mut R_0/R_g < \Delta_E < \zcut (R_g/R_0)^\beta}{\wideeq{2.5cm}}  \frac{C_i}{2\pi \alpha_s \beta_0^2} \bigg[
     \frac{\xlog(1-\lambda_c)}{1+\beta}
     -\xlog(1-\lambda_E)
     -\frac{\xlog(1-\lambda_c-(1+\beta) \lambda_g)}{1+\beta} \notag\\
     & \phantom{\wideeq{2.5cm} \frac{C_i}{2\pi \alpha_s \beta_0^2}}
     +\xlog(1-\lambda_g-\lambda_E) \bigg]\\
 & \overset{\mut< \Delta_E < \mut R_0/R_g}{\wideeq{2.5cm}}  \frac{C_i}{2\pi \alpha_s \beta_0^2} \bigg[
     \frac{\xlog(1-\lambda_c)}{1+\beta}
     -\xlog(1-\lambda_E)
     -\frac{\xlog(1-\lambda_c-(1+\beta) \lambda_g)}{1+\beta} \notag\\
     & \phantom{\wideeq{2.5cm} \frac{C_i}{2\pi \alpha_s \beta_0^2}}
     +(1-\lambda_g-\lambda_E) \log(1-\lambda_\mu)
     + (\lambda_\mu-\lambda_g-\lambda_E) \bigg]\notag\\
     & \phantom{\wideeq{2.5cm}} + \frac{C_i\as(\mu_\text{NP})}{\pi} (L_g+L_E-L_\mu)^2\\
 & \overset{\Delta_E < \mut}{\wideeq{2.5cm}}  \frac{C_i}{2\pi \alpha_s \beta_0^2} \bigg[
     \frac{\xlog(1-\lambda_c)}{1+\beta}
     -\frac{\xlog(1-\lambda_c-(1+\beta) \lambda_g)}{1+\beta}
     -\lambda_g \log(1-\lambda_\mu)
     -\lambda_g \bigg] \notag\\
     & \phantom{\wideeq{2.5cm}} + \frac{C_i\as(\mu_\text{NP})}{\pi} L_g (L_g+2 L_E-2 L_\mu).
\end{align}
Then, for $(\mut/\zcut)^{1/\beta} < R_g/R_0 < (\mut/\zcut)^{1/(1+\beta)}$, we find
\begin{align}
\bar{R}_2(\Delta_E)
 & \overset{\Delta_E > \Delta_E^{(0)}}{\wideeq{2.5cm}} \frac{C_i}{2\pi \alpha_s \beta_0^2} \bigg[
     \frac{\xlog(1-\lambda_c)}{1+\beta}
     -\xlog(1-\lambda_E)
     +\frac{\beta}{1+\beta} \xlog\left(1+\frac{\lambda_c-(1+\beta) \lambda_E}{\beta}\right) \bigg]\\
 & \overset{\zcut (R_g/R_0)^\beta < \Delta_E < \Delta_E^{(0)}}{\wideeq{2.5cm}}  \frac{C_i}{2\pi \alpha_s \beta_0^2} \bigg[
     \frac{\xlog(1-\lambda_c)}{1+\beta}
     -\xlog(1-\lambda_E)
     +\frac{\beta+\lambda_c-(1+\beta) \lambda_E}{1+\beta} \log(1-\lambda_\mu)\notag\\
     & \phantom{\wideeq{2.5cm} \frac{C_i}{2\pi \alpha_s \beta_0^2}}
     +\frac{\lambda_c+\beta \lambda_\mu-(1+\beta) \lambda_E}{1+\beta} \bigg] + \frac{C_i\as(\mu_\text{NP})}{\pi} \frac{1+\beta}{\beta} \left(L_E-\frac{L_c+\beta L_\mu}{1+\beta}\right)^2 \\
 & \overset{\mut< \Delta_E < \zcut (R_g/R_0)^\beta}{\wideeq{2.5cm}}  \frac{C_i}{2\pi \alpha_s \beta_0^2} \bigg[
     \frac{\xlog(1-\lambda_c)}{1+\beta}
     -\xlog(1-\lambda_E)
     +\frac{\beta+\lambda_c-(1+\beta)\lambda_E}{1+\beta} \log(1-\lambda_\mu)\notag\\
     & \phantom{\wideeq{2.5cm} \frac{C_i}{2\pi \alpha_s \beta_0^2}}
     +\frac{\lambda_c+\beta \lambda_\mu-(1+\beta) \lambda_E}{1+\beta} \bigg] \notag\\
     & \phantom{\wideeq{2.5cm}} + \frac{C_i\as(\mu_\text{NP})}{\pi} \left[(L_E+L_g-L_\mu)^2 - \frac{(L_c+(1+\beta) L_g-L_\mu)^2}{1+\beta}\right]\\
 & \overset{\Delta_E < \mut}{\wideeq{2.5cm}}  \frac{C_i}{2\pi \alpha_s \beta_0^2} \bigg[
     \frac{\xlog(1-\lambda_c)}{1+\beta}
     -\frac{1-\lambda_c}{1+\beta} \log(1-\lambda_\mu)
     +\frac{\lambda_c-\lambda_\mu}{1+\beta} \notag\\
     & \phantom{\wideeq{2.5cm}} + \frac{C_i\as(\mu_\text{NP})}{\pi} \left[L_g^2-\frac{(L_c+(1+\beta) L_g-L_\mu)^2}{1+\beta}+2 L_g(L_E-L_\mu) \right],
\end{align}
with $\Delta_E^{(0)}=(\zcut\mut^\beta)^{1/(1+\beta)}$. Finally, for $R_g/R_0< (\mut/\zcut)^{1/\beta}$, 
\begin{align}
\bar{R}_2(\Delta_E)
 & \overset{\Delta_E > \Delta_E^{(0)}}{\wideeq{2.5cm}} \frac{C_i}{2\pi \alpha_s \beta_0^2} \bigg[
     \frac{\xlog(1-\lambda_c)}{1+\beta}
     -\xlog(1-\lambda_E)
     +\frac{\beta}{1+\beta} \xlog\left(1+\frac{\lambda_c-(1+\beta) \lambda_E}{\beta}\right) \bigg]\\
 & \overset{\mut < \Delta_E < \Delta_E^{(0)}}{\wideeq{2.5cm}}  \frac{C_i}{2\pi \alpha_s \beta_0^2} \bigg[
     \frac{\xlog(1-\lambda_c)}{1+\beta}
     -\xlog(1-\lambda_E)
     +\frac{\beta+\lambda_c-(1+\beta) \lambda_E}{1+\beta} \log(1-\lambda_\mu)\notag\\
     & \phantom{\wideeq{2.5cm} \frac{C_i}{2\pi \alpha_s \beta_0^2}}
     +\frac{\lambda_c+\beta \lambda_\mu-(1+\beta) \lambda_E}{1+\beta} \bigg] + \frac{C_i\as(\mu_\text{NP})}{\pi} \frac{1+\beta}{\beta} \left(L_E-\frac{L_c+\beta L_\mu}{1+\beta}\right)^2 \\
 & \overset{\zcut (R_g/R_0)^\beta < \Delta_E < \mut}{\wideeq{2.5cm}}  \frac{C_i}{2\pi \alpha_s \beta_0^2} \bigg[
     \frac{\xlog(1-\lambda_c)}{1+\beta}
     -\frac{1-\lambda_c}{1+\beta} \log(1-\lambda_\mu)
     +\frac{\lambda_c-\lambda_\mu}{1+\beta} \bigg] \notag\\
     & \phantom{\wideeq{2.5cm}} + \frac{C_i\as(\mu_\text{NP})}{\pi} \left[ \frac{(L_E-L_c)^2}{\beta} - \frac{(L_\mu-L_c)^2}{1+\beta} \right]\\
 & \overset{\Delta_E < \zcut (R_g/R_0)^\beta}{\wideeq{2.5cm}}  \frac{C_i}{2\pi \alpha_s \beta_0^2} \bigg[
     \frac{\xlog(1-\lambda_c)}{1+\beta}
     -\frac{1-\lambda_c}{1+\beta} \log(1-\lambda_\mu)
     +\frac{\lambda_c-\lambda_\mu}{1+\beta} \bigg] \notag\\
     & \phantom{\wideeq{2.5cm}} + \frac{C_i\as(\mu_\text{NP})}{\pi} \left[ \beta L_g^2-\frac{(L_\mu-L_c)^2}{1+\beta} + 2 L_g (L_E-L_c-\beta L_g)\right].
\end{align}
In the above expressions, we have introduced
$L_E=\log(1/\Delta_E)$ and 
$\lambda_E=2\alpha_s \beta_0 L_E$.

\bibliography{softdrop}

\end{document}